\documentclass[12pt]{article}
\usepackage{epsfig,amsfonts,amsthm}
\usepackage{amsmath,amssymb}
\usepackage[normalem]{ulem}
\usepackage{color}
\usepackage{array}
\newcommand{\be}{\begin{equation}}
\newcommand{\ee}{\end{equation}}
\newcommand{\bea}{\begin{eqnarray}}
\newcommand{\eea}{\end{eqnarray}}

\newcommand{\doublet}[2]{ \left( \begin{array}{c}#1 \\ #2 \end{array}\right) }

\def\lsim{\mathrel{\rlap{\lower4pt\hbox{\hskip1pt$\sim$}}
    \raise1pt\hbox{$<$}}}         
\def\gsim{\mathrel{\rlap{\lower4pt\hbox{\hskip1pt$\sim$}}
    \raise1pt\hbox{$>$}}}         
    
\usepackage{amsmath}
\usepackage{amsfonts}
\usepackage{amssymb}
\usepackage{subfig}
\usepackage{wrapfig}
\usepackage{graphicx}

\def\beq{\begin{equation}}
\def\eeq{\end{equation}}
\def\bea{\begin{eqnarray}}
\def\eea{\end{eqnarray}}

\def\<{\left\langle}
\def\>{\right\rangle}

\newcommand{\bt}{\begin{tabular}}
\newcommand{\et}{\end{tabular}}

\usepackage[
top    = 3.5cm,
bottom = 3.5cm,
left   = 3.5cm,
right  = 3.5cm]{geometry}

\allowdisplaybreaks[2]
\begin{document}
\bibliographystyle{OurBibTeX}

\title{\hfill ~\\[-30mm]
                  \textbf{Dark Matter with Two Inert Doublets plus One Higgs Doublet
                }        }
\author{\\[-5mm]
Venus ~Keus\footnote{E-mail: {\tt Venus.Keus@helsinki.fi}} $^{1,2,3,4}$,\ 
Stephen ~F .~King\footnote{E-mail: {\tt King@soton.ac.uk}} $^{2}$,\ \\
Stefano ~Moretti\footnote{E-mail: {\tt S.Moretti@soton.ac.uk}}, $^{2,4}$\ 
Dorota ~Sokolowska \footnote{E-mail: {\tt Dorota.Sokolowska@fuw.edu.pl}} $^{5}$
\\ \\
  \emph{\small $^1$ Department of Physics and Helsinki Institute of Physics,}\\
  \emph{\small Gustaf Hallstromin katu 2, FIN-00014 University of Helsinki, Finland}\\
  \emph{\small $^2$ School of Physics and Astronomy, University of Southampton,}\\
  \emph{\small Southampton, SO17 1BJ, United Kingdom}\\
  \emph{\small  $^3$ Department of Physics, Royal Holloway, University of London,}\\
  \emph{\small Egham Hill, Egham TW20 0EX, United Kingdom}\\
  \emph{\small  $^4$ Particle Physics Department, Rutherford Appleton Laboratory,}\\
  \emph{\small Chilton, Didcot, Oxon OX11 0QX, United Kingdom}\\
  \emph{\small  $^5$ University of Warsaw, Faculty of Physics, Hoza 69,}\\
  \emph{\small 00-681 Warsaw, Poland}\\[4mm]}
\maketitle

\vspace*{-0.250truecm}
\begin{abstract}
\noindent
{Following the discovery of a Higgs boson, there has been renewed interest in the general 2-Higgs-Doublet Model (2HDM). A model with One Inert Doublet plus One Higgs Doublet (I(1+1)HDM), where one of the scalar doublets is ``inert'' 
(since it  has no vacuum expectation value and
does not couple to fermions) has an advantage over the 2HDM since
it provides a good Dark Matter (DM) candidate, namely the lightest inert scalar. Motivated by the existence of three fermion families, 
here we consider a model with two scalar doublets plus one Higgs doublet
(I(2+1)HDM), where  the two scalar doublets are inert.
The I(2+1)HDM has a richer phenomenology than either the I(1+1)HDM or the 2HDM. 
We discuss the new regions of  DM relic density in the I(2+1)HDM with simplified couplings
and address the possibility of constraining the model using recent results from the Large Hadron Collider (LHC) and  DM direct detection experiments.} 
\end{abstract}
 
\thispagestyle{empty}
\vfill
\newpage
\setcounter{page}{1}

\section{Introduction}

The ATLAS and CMS experiments at the Large Hadron Collider (LHC) have found evidence of a Higgs scalar with a mass of $m_h\approx 125$ GeV, which is in good agreement with earlier predictions performed using Electro-Weak (EW) precision data \cite{Aad:2012tfa,Chatrchyan:2012ufa}. Further studies are needed in order to determine whether this particle belongs to the Standard Model (SM) or to one of its extensions. However, so far, there are no reports of detection of physics Beyond the SM (BSM), neither by discovery of new particles, nor by any significant deviation from the SM prediction of the Higgs signal strengths, and strong bounds are set for the most common BSM models. 

On the other hand, new physics is expected for various theoretical and experimental reasons. One of the most important is the existence of Dark Matter (DM), \textit{stable}
on cosmological time scales, \textit{cold}, i.e., non-relativistic at the onset of galaxy formation, \textit{non-baryonic}, \textit{neutral} and \textit{weakly interacting} component of the Universe \cite{Ade:2013zuv}. Strong premises for its existence come from the galactic, cluster and horizon scales, making the modified-gravity based explanations of the observed phenomena less likely. Various candidates for such a state exist in the literature, the most well-studied being the Weakly Interacting Massive Particles (WIMPs) \cite{Jungman:1995df,Bertone:2004pz,Bergstrom:2000pn}. 

WIMP's mass may change roughly between a few
GeV and a few TeV, and the annihilation cross section is of approximately weak strength. The relic density of WIMPs is calculated with the assumption that they were in thermal equilibrium with the SM particles after inflation. Once the rate of reactions DM DM $ \leftrightarrow $ SM  SM  becomes smaller than the Hubble expansion rate of the Universe, the WIMPs freeze-out, i.e., drop out of the thermal equilibrium. After freeze-out  the co-moving WIMP density remains essentially constant, with the current value estimated by the Planck experiment to be \cite{Ade:2013zuv}:
\begin{equation}
\Omega_{DM} h^2 = 0.1199 \pm 0.0027. \label{relic}
\end{equation}

WIMPs are usually stable due to the conservation of a certain discrete symmetry. In case of the most-studied candidate in Supersymmetric (SUSY) models, neutralino (a Majorana fermion), it is the R-parity related to the imposed R-symmetry \cite{Nilles:1983ge,Haber:1984rc}. Bosonic candidates appear in the models with Universal Extra Dimensions (UED) and are made stable by the KK-parity, the remnant of momentum conservation in the extra dimension \cite{Cheng:2002ej,Servant:2002aq}. One could also consider the scalar candidates, stabilized, for example, by the conserved $Z_N$ discrete symmetry in the scalar potential, see, e.g., \cite{McDonald:1993ex,Burgess:2000yq,Deshpande:1977rw, Ma:2006km,Belanger:2012zr,Barbieri:2006dq,LopezHonorez:2006gr,Ivanov:2012hc}.

One of the simplest models that provide a scalar DM candidate is the model with One Inert Doublet plus One Higgs Doublet (I(1+1)HDM)\footnote{This  model is known in the literature as the Inert Doublet Model (IDM). We refer to it as I(1+1)HDM though for the clarification of the number of scalar doublets.}, proposed in 1976 \cite{Deshpande:1977rw}, and which has been studied extensively for the last few years (see, e.g., \cite{Ma:2006km,Barbieri:2006dq,LopezHonorez:2006gr}). In this model one $SU(2)_W$ doublet with the same quantum numbers as the SM Higgs doublet is introduced. One of the possible vacuum states in this model is $(v,0)$ where the second doublet does not develop a Vacuum Expectation Value (VEV)\footnote{The doublet that acquires a VEV is called the \textit{active} doublet and the one with no VEV is called the \textit{inert} doublet.} and therefore does not take part in the EW Symmetry Breaking (EWSB). Since this doublet does not couple to fermions, and it is by construction the only $Z_2$-odd field in the model,  it provides a stable DM candidate: the lightest state among scalar, pseudo-scalar and charged $Z_2$-odd particles.

\begin{figure}[h]
\includegraphics[scale=1]{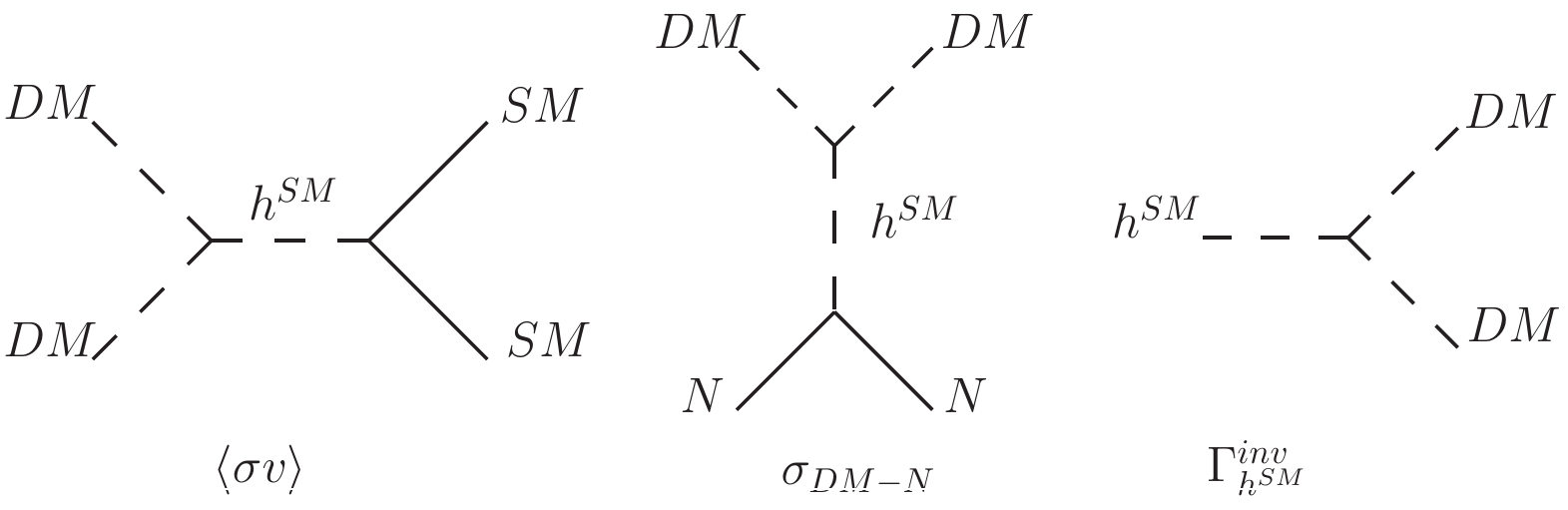}
\caption{Higgs portal Feynman diagrams; (left) DM annihilation through Higgs exchange into SM particles, (middle) DM-nucleon scattering in direct detection experiments, (right) Higgs invisible decay into two DM particles.  \label{higgsportal}}
\end{figure}

The I(1+1)HDM can be treated as an example of the Higgs-portal type of DM model, where the DM sector communicates with the SM sector through the Higgs boson exchange \cite{Patt:2006fw,Chu:2011be,Queiroz:2014yna}. As a result, the DM-Higgs coupling, $g_{DMh}$, governs the DM annihilation rate $\langle\sigma v\rangle$, the DM-nucleon scattering cross-section $\sigma_{DM-N}$ and the Higgs invisible decays (see Fig.(\ref{higgsportal})). Normally, fulfilling current experimental constraints for these three types of processes at the same time is a very difficult task, as shown for e.g. in \cite{Mambrini:2011ik,Djouadi:2011aa,Djouadi:2012zc}. A possible solution to this problem is destroying
the simple relation between the annihilation rate and the direct detection cross-section by introducing coannihilation processes, between DM  and other inert particles, which are close in mass. Coannihilation processes lead to an increase or decrease of the effective annihilation cross-section, which in turn gives respectively smaller or larger DM relic density values. In the I(1+1)HDM, for example, the DM candidate could coannihilate with neutral and/or charged $Z_2$-odd particles. In models with a richer particle spectrum, more coannihilation processes could come to play.

One could simply extend the I(1+1)HDM by introducing an extra \textit{inert} $SU(2)_W$ doublet with the same quantum numbers as the SM Higgs doublet, resulting in a 3-Higgs-Doublet Model (3HDM). As the next simplest example beyond 2HDMs, which has been extensively studied in the literature, 3HDMs are very well motivated. Furthermore, all possible finite symmetries in 3HDMs have recently been identified \cite{Ivanov:2012fp}. 

3HDMs may address the problem of the origin and nature of the three fermion families. Indeed it is possible that the symmetry of the three Higgs doublets could describe the symmetry of the three families of quarks and leptons. In a recent paper \cite{Keus:2013hya}, we studied symmetric 3HDMs and derived the conditions under which the vacuum alignments $(0,0,v_3)$, $(0,v_2,v_3)$ and $(v_1,v_2,v_3)$ are minima of the potential. Here we focus on the alignment $(0,0,v_3)$, which is of particular interest because of its I(1+1)HDM similarity and the absence of Flavour Changing Neutral Currents (FCNCs)\footnote{A 3HDM with $(0,v_2,v_3)$ vacuum alignment has been considered in \cite{Grzadkowski:2010au} wherein it was termed IDM2. Using our nomenclature, this model may be referred to as the I(1+2)HDM.}.

In this paper, we study a model with Two Inert Doublets plus One Higgs Doublet (I(2+1)HDM). The two inert doublets are $Z_2$-odd and the active doublet is $Z_2$-even which plays the role of the SM Higgs doublet. The I(2+1)HDM may be regarded as an extension to the I(1+1)HDM. In this scenario the $Z_2$-odd particle content is doubled with respect to the I(1+1)HDM, and so new possibilities of DM (co)annihilation appear. One can have up to six (co)annihilating states, which introduce a very different behaviour with respect to models with fewer number of states in the inert sector.

The layout of the paper is as follows. In section \ref{construction} we construct the $Z_2$-symmetric I(2+1)HDM potential and study the $(0,0,v_3)$ vacuum point. In section \ref{constraints} theoretical and experimental constraints on the parameters of the model are derived and presented. In section \ref{annihilation} we list all phenomenologically viable DM (co)annihilation scenarios in our model. In section \ref{constrained-3IDM} we study in detail a simplified version of the I(2+1)HDM with reduced number of parameters  and present relic density plots in different scenarios. We finally draw our conclusions in section \ref{conclusion}. The Feynman rules are presented in the Appendix.

\section{Constructing the I(2+1)HDM potential}\label{construction}
In a general N-Higgs-Doublet Model (NHDM), the scalar potential which is symmetric under a group $G$ of phase rotations can be written as 
\be 
V = V_0 + V_G
\ee
where $V_0$ is invariant under any phase rotation and $V_G$ is a collection of extra terms ensuring the symmetry group $G$ \cite{Ivanov:2011ae}.
The most general phase invariant part of the I(2+1)HDM potential has the following form:
\bea
\label{V0-3HDM}
V_0 &=& - \mu^2_{1} (\phi_1^\dagger \phi_1) -\mu^2_2 (\phi_2^\dagger \phi_2) - \mu^2_3(\phi_3^\dagger \phi_3) \\
&&+ \lambda_{11} (\phi_1^\dagger \phi_1)^2+ \lambda_{22} (\phi_2^\dagger \phi_2)^2  + \lambda_{33} (\phi_3^\dagger \phi_3)^2 \nonumber\\
&& + \lambda_{12}  (\phi_1^\dagger \phi_1)(\phi_2^\dagger \phi_2)
 + \lambda_{23}  (\phi_2^\dagger \phi_2)(\phi_3^\dagger \phi_3) + \lambda_{31} (\phi_3^\dagger \phi_3)(\phi_1^\dagger \phi_1) \nonumber\\
&& + \lambda'_{12} (\phi_1^\dagger \phi_2)(\phi_2^\dagger \phi_1) 
 + \lambda'_{23} (\phi_2^\dagger \phi_3)(\phi_3^\dagger \phi_2) + \lambda'_{31} (\phi_3^\dagger \phi_1)(\phi_1^\dagger \phi_3).  \nonumber
\eea
Constructing the $Z_2$-symmetric part of the potential depends on the generator of the group. The $Z_2$ generator which forbids FCNCs and is respected by the vacuum alignment $(0,0,v)$ has the following form
\be 
g=  \mathrm{diag}\left(-1, -1, 1 \right). 
\ee
The terms ensuring the $Z_2$ group generated by $g$ are
\be 
V_{Z_2} = -\mu^2_{12}(\phi_1^\dagger\phi_2)+  \lambda_{1}(\phi_1^\dagger\phi_2)^2 + \lambda_2(\phi_2^\dagger\phi_3)^2 + \lambda_3(\phi_3^\dagger\phi_1)^2  + h.c. 
\label{Z_2-3HDM}
\ee
which need to be added to $V_0$ in Eq.~(\ref{V0-3HDM}) to result in an I(2+1)HDM potential which is only $Z_2$-symmetric.
We shall not consider CP-violation in this paper, therefore, we require all parameters of the potential to be real.

\subsection{Mass eigenstates}

We define the doublets as
\be 
\phi_1= \doublet{$\begin{scriptsize}$ H^+_1 $\end{scriptsize}$}{\frac{H^0_1+iA^0_1}{\sqrt{2}}},\quad 
\phi_2= \doublet{$\begin{scriptsize}$ H^+_2 $\end{scriptsize}$}{\frac{H^0_2+iA^0_2}{\sqrt{2}}}, \quad 
\phi_3= \doublet{$\begin{scriptsize}$ G^+ $\end{scriptsize}$}{\frac{v+h+iG^0}{\sqrt{2}}}, 
\label{explicit-fields}
\ee
with two inert doublets ($\phi_1$ and $\phi_2$) and one active doublet ($\phi_3$) where the latter plays the role of the SM Higgs doublet, with $h$ being the SM-Higgs boson.
The CP-even/odd neutral $Z_2$-odd fields from the inert doublets could in principle be DM candidates since only the fields from the active doublet couple to the fermions. To stabilise the DM candidate from decaying into SM particles, we make use of the conserved $Z_2$ symmetry of the potential after EWSB. 
To make sure that the entire Lagrangian and not only the scalar potential is $Z_2$ symmetric, we assign an even $Z_2$ parity to all SM particles, identical to the $Z_2$ parity of the only doublet that couples to them, i.e., the active doublet $\phi_3$. With this parity assignment FCNCs are avoided as the extra doublets are forbidden to decay to fermions by $Z_2$ conservation.

Note that the Yukawa Lagrangian in this model is identical to the SM one, with $\phi_3$ playing the role of the SM Higgs doublet:
\bea 
\mathcal{L}_{Yuk} &=& \Gamma^u_{mn} \bar{q}_{m,L} \tilde{\phi}_3 u_{n,R} + \Gamma^d_{mn} \bar{q}_{m,L} \phi_3 d_{n,R} \nonumber\\
&& +  \Gamma^e_{mn} \bar{l}_{m,L} \phi_3 e_{n,R} + \Gamma^{\nu}_{mn} \bar{l}_{m,L} \tilde{\phi}_3 {\nu}_{n,R} + h.c.  
\eea
 
The point $(0,0,\frac{v}{\sqrt{2}})$ becomes the minimum of the potential at 
\be
v^2= \frac{\mu^2_3}{\lambda_{33}} 
\ee

Expanding the potential around this vacuum point results in the mass spectrum below, where the pairs of inert scalar/pseudo-scalar/charged base fields ($H^0_{1,2}, A^0_{1,2}, H^\pm_{1,2}$) are rotated by:
\be 
R_{\theta_i}= 
\left( \begin{array}{cc}
\cos \theta_i & \sin \theta_i \\
-\sin \theta_i & \cos \theta_i\\
\end{array} \right).
\ee
into the mass eigenstates identified in boldface fonts.
\begin{footnotesize}
\bea
&& \textbf{G}^0  : \quad m^2_{G^0}=0 \nonumber\\
&& \textbf{G}^\pm  : \quad m^2_{G^\pm}=0 \nonumber\\
&& \textbf{h} : \quad m^2_{h}= 2\mu_3^2 \nonumber\\
&& \textbf{H}_1 = \cos\theta_h H^0_{1}+ \sin\theta_hH^0_{2} : \quad m^2_{H_1}=  (-\mu^2_1 + \Lambda_{\phi_1})\cos^2\theta_h + (- \mu^2_2 + \Lambda_{\phi_2}) \sin^2\theta_h -2\mu^2_{12} \sin\theta_h \cos\theta_h \nonumber\\
&& \textbf{H}_2 = -\sin\theta_h H^0_{1}+ \cos\theta_hH^0_{2} : \quad m^2_{H_2}=  (-\mu^2_1 + \Lambda_{\phi_1})\sin^2\theta_h + (- \mu^2_2 + \Lambda_{\phi_2}) \cos^2\theta_h + 2\mu^2_{12} \sin\theta_h \cos\theta_h \nonumber\\
&& \qquad \qquad  \mbox{where} \quad \Lambda_{\phi_1}= \frac{1}{2}(\lambda_{31} + \lambda'_{31} +  2\lambda_3)v^2  \nonumber\\
&& \qquad \qquad \qquad \qquad \Lambda_{\phi_2}= \frac{1}{2}(\lambda_{23} + \lambda'_{23} +2\lambda_2 )v^2   \nonumber\\
&& \qquad \qquad \qquad \qquad \tan 2\theta_h = \frac{2\mu^2_{12}}{\mu^2_1 -\Lambda_{\phi_1} - \mu^2_2 + \Lambda_{\phi_2}} \nonumber\\[2mm]
&& \textbf{H}^\pm_1 = \cos\theta_cH^\pm_{1}+ \sin\theta_cH^\pm_{2} : \quad m^2_{H^\pm_1}=  (-\mu^2_1 + \Lambda'_{\phi_1})\cos^2\theta_c + (- \mu^2_2 + \Lambda'_{\phi_2}) \sin^2\theta_c -2\mu^2_{12} \sin\theta_c \cos\theta_c \nonumber\\
&& \textbf{H}^\pm_2 = -\sin\theta_cH^\pm_{1}+ \cos\theta_cH^\pm_{2} :\quad m^2_{H^\pm_1}= (-\mu^2_1 + \Lambda'_{\phi_1})\sin^2\theta_c + (- \mu^2_2 + \Lambda'_{\phi_2}) \cos^2\theta_c + 2\mu^2_{12} \sin\theta_c \cos\theta_c \nonumber\\
&& \qquad \qquad  \mbox{where} \quad \Lambda'_{\phi_1}= \frac{1}{2}(\lambda_{31})v^2  \nonumber\\
&& \qquad \qquad \qquad \qquad  \Lambda'_{\phi_2}= \frac{1}{2}(\lambda_{23} )v^2   \nonumber\\
&& \qquad \qquad \qquad \qquad \tan 2\theta_c = \frac{2\mu^2_{12}}{\mu^2_1 - \Lambda'_{\phi_1} - \mu^2_2 + \Lambda'_{\phi_2}} \nonumber\\[2mm]
&& \textbf{A}_1 = \cos\theta_aA^0_{1}+ \sin\theta_a A^0_{2}: \quad m^2_{A_1}= (-\mu^2_1 + \Lambda''_{\phi_1})\cos^2\theta_a + (- \mu^2_2 + \Lambda''_{\phi_2}) \sin^2\theta_a -2\mu^2_{12} \sin\theta_a \cos\theta_a \nonumber\\
&& \textbf{A}_2 = -\sin\theta_aA^0_{1}+ \cos\theta_a A^0_{2} : \quad m^2_{A_2}= (-\mu^2_1 + \Lambda''_{\phi_1})\sin^2\theta_a + (- \mu^2_2 + \Lambda''_{\phi_2}) \cos^2\theta_a + 2\mu^2_{12} \sin\theta_a \cos\theta_a \nonumber\\
&& \qquad \qquad  \mbox{where} \quad \Lambda''_{\phi_1}= \frac{1}{2}(\lambda_{31} + \lambda'_{31} - 2\lambda_3)v^2  \nonumber\\
&& \qquad \qquad \qquad \qquad  \Lambda''_{\phi_2}= \frac{1}{2}(\lambda_{23} + \lambda'_{23} -2\lambda_2 )v^2   \nonumber\\
&& \qquad \qquad \qquad \qquad \tan 2\theta_a = \frac{2\mu^2_{12}}{\mu^2_1 - \Lambda''_{\phi_1} - \mu^2_2 + \Lambda''_{\phi_2}} \nonumber
\eea
\end{footnotesize}

There are two generations of physical inert states; fields from the first generation, $(H_1,A_1,H^\pm_1)$ are chosen to be lighter than the respective fields from the second generation, $(H_2,A_2,H^\pm_2)$, with $H_1$ being the lightest of them all, i.e., a DM candidate:
\begin{equation}
m_{H_1} < m_{H_2}, m_{A_{1,2}},m_{H^\pm_{1,2}}.
\end{equation}

\begin{figure}[ht!]
\centering
\includegraphics[scale=0.8]{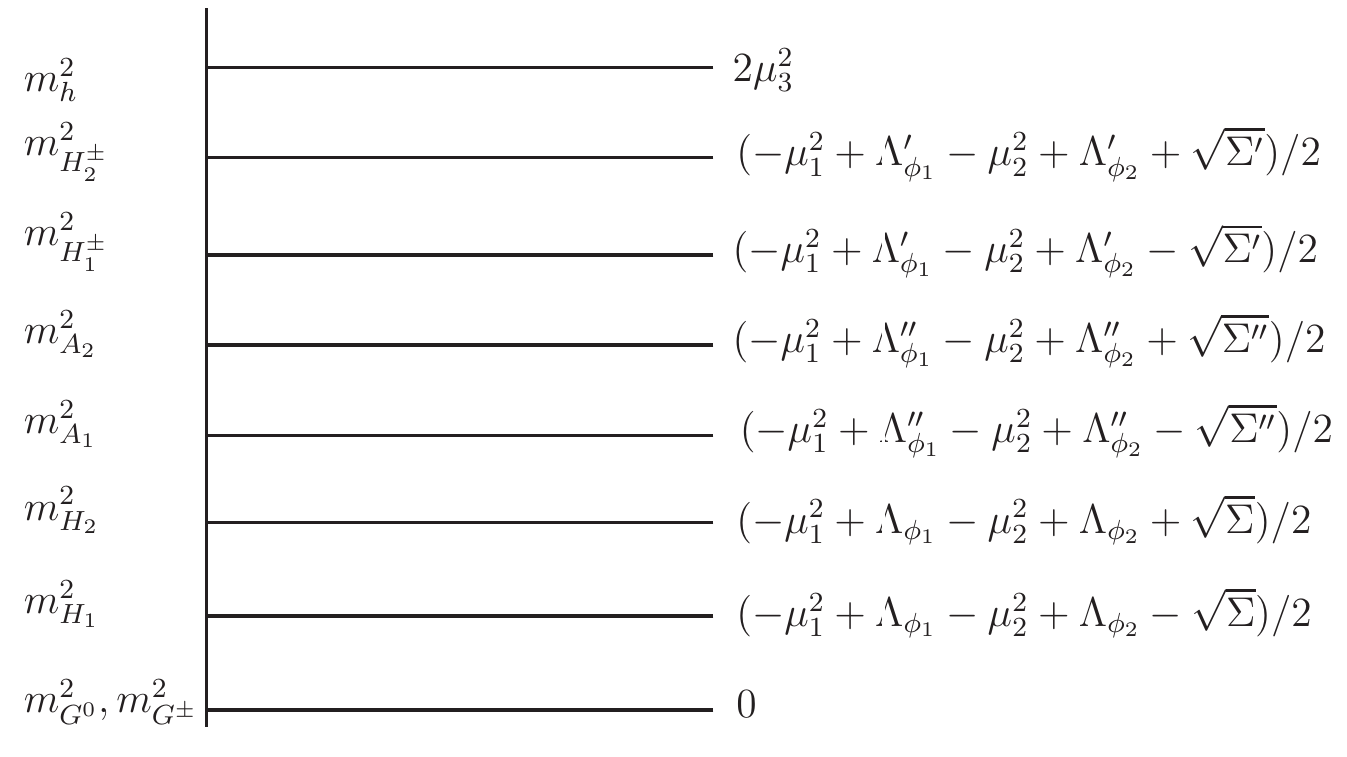} 
\caption{Schematic mass-squared spectrum of the $Z_2$ symmetric I(2+1)HDM, where 
$\Sigma= 4\mu^4_{12} + (\mu^2_1-\Lambda_{\phi_1} -\mu^2_2 +\Lambda_{\phi_2})^2 $, 
$\Sigma' = 4\mu^4_{12} + (\mu^2_1-\Lambda'_{\phi_1} -\mu^2_2 +\Lambda'_{\phi_2})^2 $ and 
$\Sigma''=4\mu^4_{12} + (\mu^2_1-\Lambda''_{\phi_1} -\mu^2_2 +\Lambda''_{\phi_2})^2 $.}
\label{Masses-fig}
\end{figure}

The mass spectrum has the schematic form shown in Fig.(\ref{Masses-fig}), provided the CP-even neutral inert particles are lighter than the CP-odd and charged inert particles, which puts the following constraints on the parameters:
\be 
2\lambda_2 , 2\lambda_3 < \lambda'_{23}, \lambda'_{31} < 0. 
\label{lambda-assumption} 
\ee
We also consider cases where the mass alignment is changed, but where $H_1$ is always the lightest inert state. In the remainder of the paper the notations $H_1$ and $DM$ will be used interchangeably.

\section{Constraints on parameters}\label{constraints}

The parameters of the potential can be divided into the following categories:
\begin{itemize}
\item $\mu_3, \lambda_{33}$ are Higgs field parameters, given by the Higgs mass. We use the value $125$ GeV for the latter, so that \cite{Aad:2014aba,CMS:2014ega} 
\begin{equation} m^2_h = 2\mu^2_3 = 2\lambda_{33} v^2.\end{equation}

\item $\mu_{1},\mu_{2},\mu_{12}, \lambda_{31},\lambda_{23},\lambda'_{31},\lambda'_{23},\lambda_{2}, \lambda_{3}$ are related to masses of inert scalars and their couplings with the visible sector (through $h$). These 9 parameters can in principle be determined by independent masses, mixing angles or couplings
and the ranges that we allow for them in our numerical studies are
\bea
&& -10~ \mbox{ TeV}^2 < \mu^2_{1},\mu^2_{2},\mu^2_{12} < 10~ \mbox {TeV}^2 \\
&& -0.5 < \lambda_{31},\lambda_{23},\lambda'_{31},\lambda'_{23},\lambda_{2}, \lambda_{3} < 0.5 \nonumber
\eea

\item $\lambda_{11},\lambda_{22},\lambda_{12}, \lambda'_{12}$ are  inert sector  parameters (inert scalars self-interactions), so that relic density calculations do not depend on these and therefore DM measurements do not constrain them\footnote{Any bound on these parameters should then come from collider limits.}, for which we scan the ranges
\be 
 0 < \lambda_{11},\lambda_{22},\lambda_{12}, \lambda'_{12} < 0.5.
\ee

\end{itemize}

\subsection{Theoretical constraints}
Theoretical requirements of positivity of mass eigenstates, bounded-ness of the potential and positive-definite-ness of the Hessian put the following constraints on the potential.
\begin{enumerate}
\item 
\textbf{Positivity of the mass eigenstates}
\bea
\label{positivity}
&& \bullet \qquad \mu^2_3 > 0    \\
&& \bullet \quad -2\mu^2_1 + \lambda_{31}v^2> 0 \nonumber\\
&& \bullet \quad  -2\mu^2_1 + (\lambda_{31}+\lambda'_{31} )v^2> 0 \nonumber\\
&& \bullet \quad -2\mu^2_1 + (\lambda_{31}+\lambda'_{31} -2\lambda_3 )v^2> 0   \nonumber\\
&& \bullet \quad -2\mu^2_2 + \lambda_{23}v^2> 0 \nonumber\\
&& \bullet \quad -2\mu^2_2 + (\lambda_{23}+\lambda'_{23} )v^2> 0 \nonumber\\
&& \bullet \quad -2\mu^2_2 + (\lambda_{23}+\lambda'_{23} -2\lambda_2 )v^2> 0   \nonumber\\
&& \bullet \quad  -2\mu^2_1 -2\mu^2_2  + (\lambda_{31}+ \lambda_{23})v^2> 4|\mu^2_{12}|   \nonumber\\
&& \bullet \quad -2\mu^2_1 -2\mu^2_2  + (\lambda_{31}+ \lambda_{23} +\lambda'_{31}+ \lambda'_{23} )v^2> 4|\mu^2_{12}|   \nonumber\\
&& \bullet \quad -2\mu^2_1 -2\mu^2_2  + (\lambda_{31}+ \lambda_{23} +\lambda'_{31}+ \lambda'_{23} - 2\lambda_3 -2\lambda_2)v^2> 4|\mu^2_{12}|   \nonumber
\eea

\item
\textbf{Bounded-ness of the potential}
\\For the $V_0$ part of the potential to have a stable vacuum (bounded from below) the following conditions are required\footnote{These conditions are resulted from requiring the quartic part of the potential to be positive as the fields $\phi_i \to \infty$. The ``copositivity" method suggested in \cite{Kannike:2012pe} will result in less restrictive constrains.}
\bea
&& \bullet\quad \lambda_{11}, \lambda_{22}, \lambda_{33} > 0 \\
&& \bullet\quad \lambda_{12} + \lambda'_{12} > -2 \sqrt{\lambda_{11}\lambda_{22}} \nonumber\\
&& \bullet\quad \lambda_{23} + \lambda'_{23} > -2 \sqrt{\lambda_{22}\lambda_{33}} \nonumber\\
&& \bullet\quad \lambda_{31} + \lambda'_{31} > -2 \sqrt{\lambda_{33}\lambda_{11}} \nonumber
\eea
We also require the parameters of the $V_{Z_2}$ part to be smaller than the parameters of the $V_0$ part:
\be 
\bullet\quad |\lambda_2|, |\lambda_2|, |\lambda_3| < |\lambda_{ii}|, |\lambda_{ij}|, |\lambda'_{ij}| , \quad i\neq j = 1,2,3.
\ee

\item
\textbf{Positive-definite-ness of the Hessian}
\\For the point $(0,0,\frac{v}{\sqrt{2}})$ to be a minimum of the potential, the second order derivative matrix must have positive definite determinant. Therefore, the following constraints are required:
\bea 
&& \bullet\qquad   \mu^2_3  > 0   \\
&& \bullet\quad  -2\mu^2_2  + (\lambda_{23}+\lambda'_{23} )v^2   >0  \nonumber\\
&& \bullet\quad  -2\mu^2_1  + (\lambda_{31}+\lambda'_{31} )v^2   >0  \nonumber\\
&& \bullet\quad  \biggl(-2\mu^2_1  +(\lambda_{31}+\lambda'_{31} )v^2 \biggr) \biggl( -2\mu^2_2  +(\lambda_{23}+\lambda'_{23} )v^2 \biggr) > 4\mu^4_{12}  \nonumber
\eea

\end{enumerate}

\subsection{Experimental constraints}

Relevant constraints limit the parameters from different experiments.
\subsubsection{Collider constraints}

\begin{itemize}
\item \textbf{LEP limits}\\
Measurements done at LEP limit the invisible decays of $Z$ and $W^\pm$ gauge bosons, require that  \cite{Cao:2007rm,Lundstrom:2008ai}
\bea 
&& \bullet\quad m_{H_i^\pm} + m_{H_i,A_i} > m_{W^\pm} \\
&& \bullet\quad m_{H_i} + m_{A_i} > m_Z \nonumber\\
&& \bullet\quad 2m_{H_i^\pm} > m_Z \nonumber
\eea

Also, LEP provides a model-independent lower limit for the mass of the charged scalars: 
\be 
\bullet\quad  m_{H^\pm_i} > 70-90 \quad \mbox{GeV}.
\ee
Searches for charginos and neutralinos at LEP have been translated into limits of region of masses in the I(1+1)HDM \cite{Lundstrom:2008ai} where for 
\be  
m_H < 80 \quad \mbox{GeV} \quad \mbox{and} \quad m_A < 100 \quad \mbox{GeV} \nonumber
\ee
the following region is excluded
\be 
\bullet\quad m_A - m_H > 8 \quad \mbox{GeV}.  
\ee
We have taken this limit into account in our numerical studies for any pair of CP-even and CP-odd particles.

\item \textbf{LHC limits}\\
Measurements of invisible Higgs decays limit models in which the Higgs boson can decay into lighter particles which escape detection. Current experimental values provided by the ATLAS and CMS experiments are \cite{atlasbr,Chatrchyan:2014tja}
on the ensuing Branching ratio ($Br$) are:
\be
Br(h \to inv.) < 37 \%,
\ee
\be
Br(h \to inv.) < 58 \%.
\ee
where $h \to inv.$ represents the SM-Higgs decay to invisible particles channel which in our case is the $h \to H_1H_1$ channel.
Global fits on Higgs signal strengths require the invisible $Br$ of a Higgs boson with SM couplings but additional invisible decay modes to be limited to \cite{Belanger:2013xza}
\be 
\bullet\quad Br(h \rightarrow  inv.) < 23\%  \quad \mbox{at} \quad  95\% \mbox{ CL} \label{BR-inv-global}
\ee

The invisible Higgs decay into two scalar particles, CP-even or CP-odd, denoted by $S$ is given by
\be 
\Gamma(h\to SS) =\frac{\lambda^2 v^2}{32 \pi m_h} \sqrt{1-\frac{4 m_S^2}{m_h^2}} \qquad \mbox{with} \quad S=H_1, H_2, A_1, A_2,
\ee
where $m_S<m_h/2$ is the scalar particle mass and $\lambda$ is its coupling to the SM Higgs boson. In first approximation (with only one scalar, say $H_1$, having a mass below $m_h/2 \approx 62.5$ GeV) the invisible decay rate is such that
\be
Br(h\to inv.) = \frac{\Gamma(h\to H_1H_1)}{\Gamma_h^{\rm SM}+\Gamma(h \to H_1H_1)}.
\ee
The limit from Eq.~(\ref{BR-inv-global}) leads to strong constraints on the $H_1H_1h$ coupling (roughly $\lambda \lesssim 0.02$ for masses $m_{H_1} \lesssim m_h/2$). In general, this will lead to tension between the LHC limits, which favour smaller couplings, and the relic density limits (discussed below), which favour larger couplings needed for effective DM annihilation.

Note that the presence of additional charged scalar states, $H^\pm_{1,2}$, may modify the Higgs diphoton decay channel and lead to deviation from the SM value defined as\footnote{In the I(2+1)HDM the main production channel is through gluon fusion and the Higgs boson is SM-like, which leads to  $\sigma(gg\to h)^{\textrm{I(2+1)HDM}} = \sigma(gg\to h)^{\textrm{SM}}$ and thus to the simplification in Eq.~(\ref{rgg}).}:
\be 
\label{rgg}
\mu_{\gamma \gamma}:=\frac{\sigma(pp\to h\to \gamma\gamma)^{\textrm{I(2+1)HDM}}}{\sigma(pp\to h\to \gamma\gamma)^{\textrm  {SM}}}
\approx \frac{\Gamma(h\to \gamma\gamma)^{\mathrm{I(2+1)HDM}}}{\Gamma(h\to \gamma\gamma)^{\mathrm{SM}}}\frac{\Gamma(h)^{\mathrm{SM}}}{\Gamma(h)^{\mathrm{I(2+1)HDM}}} \, ,
\ee
where $\Gamma(h)^{\mathrm{SM}}$ and $\Gamma(h)^{\mathrm{I(2+1)HDM}}$ are the total decay widths of the Higgs boson in the SM and the I(2+1)HDM, respectively, while $\Gamma(h\to \gamma\gamma)^{\mathrm{SM}}$ and $\Gamma(h\to \gamma\gamma)^{\mathrm{I(2+1)HDM}}$ are the respective partial decay widths for the process $h\to\gamma\gamma$.

Currently, experimental values provided by the CMS and ATLAS collaborations are in agreement with the SM prediction $\mu_{\gamma \gamma}=1$ within the experimental errors \cite{Aad:2014aba,Khachatryan:2014ira}
\bea
\textrm{ATLAS} &: & \mu_{\gamma \gamma}=1.29 \pm 0.30, \label{rg_atlas} \nonumber\\
\textrm{CMS} & : &  \mu_{\gamma \gamma}=1.14^{+0.26}_{-0.23}. \label{rg_cms}
\eea
However, if the future combined value with reduced uncertainties shows significant deviation from $ \mu_{\gamma \gamma} = 1$, it will provide strong constraints for  multi-scalar models. 

\end{itemize}

\subsubsection{Dark Matter constraints}

\begin{itemize}
\item \textbf{Relic density constraints}\\
DM relic density $\Omega_{DM} h^2$ is constrained by the combined Planck and WMAP results to be \cite{Ade:2013zuv}:
\be 
\Omega_{DM} h^2 = 0.1199 \pm 0.0027,
\ee
which leads to the 3$\sigma$ bound:
\be 
\bullet \quad  0.1118<\Omega_{DM}h^2 <0.128. \label{PLANCK_lim}
\ee
If a DM candidate fulfils this requirement, then it constitutes 100\% of the DM in the Universe. However, a subdominant DM candidate is allowed if its relic density is smaller than $0.1118$. Regions of the parameter space corresponding to $\Omega_{DM}h^2$ larger than the Planck upper limit are excluded.

In our work we use the micrOMEGAs 3.5 package to compute the relic density \cite{Belanger:2013oya}. All annihilation and coannihilation channels are taken into account, including final states with one or two virtual gauge bosons in all cases relevant for the chosen values of masses.

\item \textbf{Direct detection constraints}\\
Neutral and non-relativistic WIMPs are expected to interact mainly with the atomic nuclei, whose nuclear recoil energy is to be measured by the DM detector. The current strongest upper limit on the Spin Independent (SI) scattering cross section $\sigma_{DM-N}$ is provided by the LUX experiment \cite{Akerib:2013tjd}:
\be  
\bullet \quad \sigma_{DM-N} < 7.6 \times 10^{-46}~{\rm cm}^2 \quad \mbox{for} \quad m_{\rm DM}= 33 \mbox{ GeV}. 
\ee
Limits from XENON100 (2012) are slightly weaker \cite{Aprile:2012nq}, with the strongest exclusion limit 
\be 
\bullet \quad  \sigma_{DM-N} < 2 \times 10^{-45}~{\rm cm}^2\quad \mbox{for} \quad m_{\rm DM}= 55 \mbox{ GeV}.
\ee

\item \textbf{Indirect detection constraints}\\
The indirect evidence for DM can be provided by measurements of the excess in the cosmic ray fluxes coming from the annihilation of DM in the Milky Way halo. 
The strongest constraints for light DM\footnote{PAMELA and AMS provide strong constraints for models in which DM
annihilates predominantly into $e^+e^-$ or $\tau^+\tau^-$ pairs, which is not the case of the I(2+1)HDM considered in this work.} annihilating into $b\bar b$ or $\tau^+\tau^-$ is provided by the measurements of the gamma-ray flux from Dwarf Spheroidal Galaxies by the Fermi-LAT satellite, ruling out the canonical cross-section \cite{Ackermann:2013yva,Ackermann:2011wa}:
\be  
\bullet \quad \langle \sigma v\rangle \approx 3\times 10^{-26}~{\rm cm}^3/{\rm s} \quad \mbox{for} \quad m_{\rm DM} \lesssim 25-40 \mbox{ GeV}.
\ee  
For the heavier DM candidates PAMELA and Fermi-LAT experiments provide similar limits of 
\be 
\bullet \quad \langle \sigma v\rangle \approx 10^{-25}~{\rm cm}^3/{\rm s} \quad \mbox{for} \quad m_{\rm DM}=200 \mbox{ GeV} 
\ee
in the $b\bar b,\tau^+\tau^-$ or $W^+W^-$ channels \cite{Cirelli:2013hv}. HESS measurements of signals coming from the Galactic Centre set limits of $ \langle \sigma v\rangle \approx 10^{-25}-10^{-24}~{\rm cm}^3/{\rm s}$ for DM
masses up to TeV scales \cite{Abramowski:2011hc}.

\end{itemize}

\section{DM (co)annihilation in the I(2+1)HDM}\label{annihilation}
The relic density of the scalar DM candidate, $S$, after freeze-out is given by the solution of the Boltzmann equation:
\begin{equation}
\frac{d n_S}{dt} = - 3 H n_S - \langle \sigma_{eff} v \rangle (n_S^2 - n^{eq \; 2}_{S}),
\end{equation}
where the thermally averaged effective (co)annihilation cross-section contains all relevant annihilation processes of any $S_i S_j$ pair into SM particles:
\begin{equation}
\langle \sigma_{eff} v \rangle = \sum_{ij} \langle \sigma_{ij} v_{ij} \rangle \frac{n^{eq}_i}{n^{eq}_S} \frac{n^{eq}_j}{n^{eq}_S},
\end{equation}
where
\begin{equation}
\frac{n^{eq}_i}{n^{eq}_S} \sim \exp({-\frac{m_i - m_S}{T}}).
\end{equation}
Therefore, only processes for which the mass splitting between a state $S_i$ and the lightest $Z_2$-odd particle $S$ are comparable to the thermal bath temperature $T$ provide a sizeable contribution to this sum.

\begin{figure}[h!]
\centering
\includegraphics[scale=0.7]{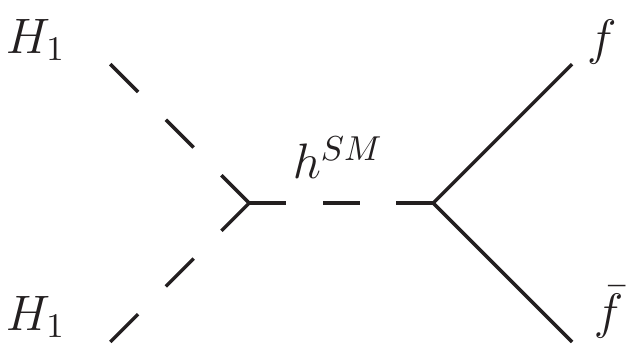}
\caption{The most important channel for the DM annihilation in a large region of parameter space, where $H_1$ is the DM candidate in our model and $h^{\rm SM}$ represents the SM-Higgs boson.}
\label{annihilation-1}
\end{figure}

The I(2+1)HDM studied here shares many features of a Higgs-portal DM model. In a large region of parameter space the most important channel for the DM annihilation is shown in Fig.(\ref{annihilation-1}).
The efficiency of this annihilation channel depends both on the value of DM mass and its coupling to the Higgs particle. In general, if $m_{\rm DM} < m_h/2$, then one needs a coupling that is relatively large to produce relic density in agreement with Eq.~(\ref{relic}). In this case a small DM-Higgs coupling results in too large a relic density and leads to the overclosure of the Universe. 
Note that a relic density below the Planck value does not exclude the DM candidate in the model, but requires another component to the DM to complete the deficit in the relic density.

\begin{figure}[h!]
\centering
\includegraphics[scale=0.7]{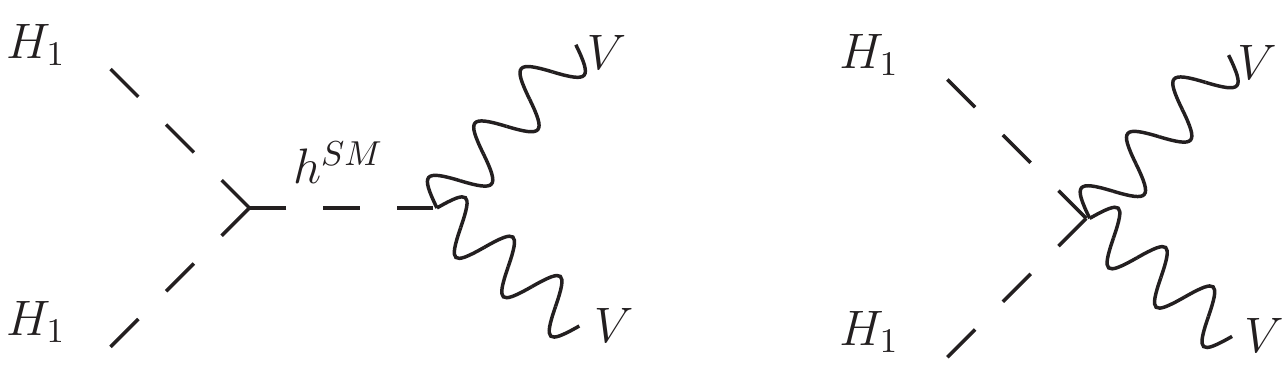}
\caption{Diagrams contributing to the total annihilation cross section when $m_{\rm DM} > m_W$, where $V$ is any of SM gauge bosons.}
\label{annihilation-2}
\end{figure}

\begin{figure}[h!]
\centering
\includegraphics[scale=0.7]{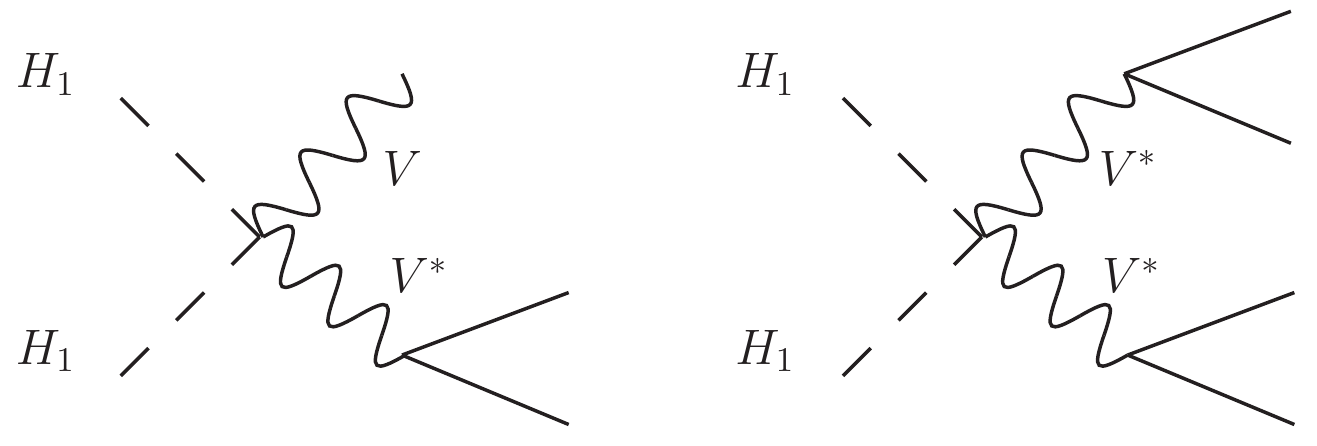}
\caption{Diagrams with off-shell gauge bosons which could play an important role in the $m_{\rm DM} < m_W$ region.}
\label{annihilation-3}
\end{figure}

The diagrams shown in Fig.(\ref{annihilation-2}) also contribute to the total annihilation cross section, where $V$ is any of SM gauge bosons.
Contribution from these diagrams is suppressed when the DM mass is smaller than $m_W$, however, as studies have shown, diagrams with off-shell gauge bosons may be very important for $m_{\rm DM} < m_W$ in models such as the I(1+1)HDM. In our analysis the diagrams shown in Fig.(\ref{annihilation-3}) are also included.

\begin{figure}[h!]
\centering
\includegraphics[scale=0.9]{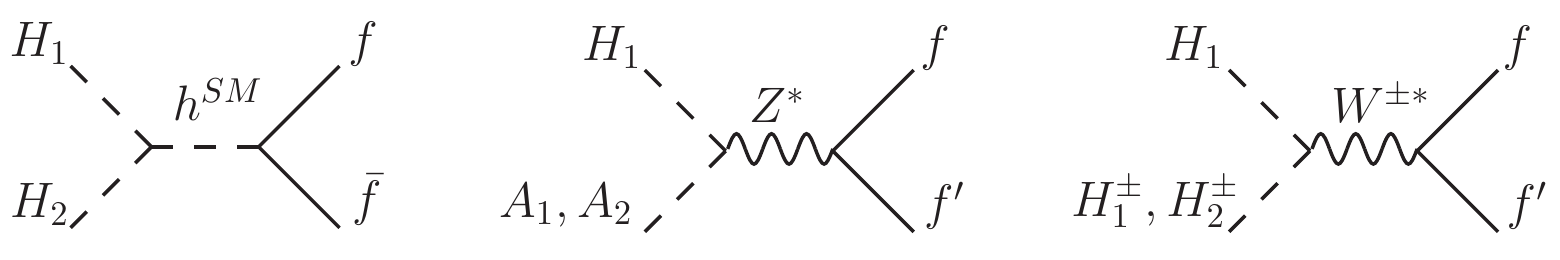}
\caption{The coannihilation channels appearing in our model.}
\label{annihilation-4}
\end{figure}

Coannihilation effects play an important role in scenarios with multiple particles that are close in mass. Particles up to 20\% heavier than the DM candidate may influence the DM relic density. Therefore, the coannihilation diagrams should be included in calculating the effective annihilation cross section. The coannihilation channels shown in Fig.(\ref{annihilation-4}) appear in our studies.

\begin{figure}[h!]
\centering
\includegraphics[scale=0.7]{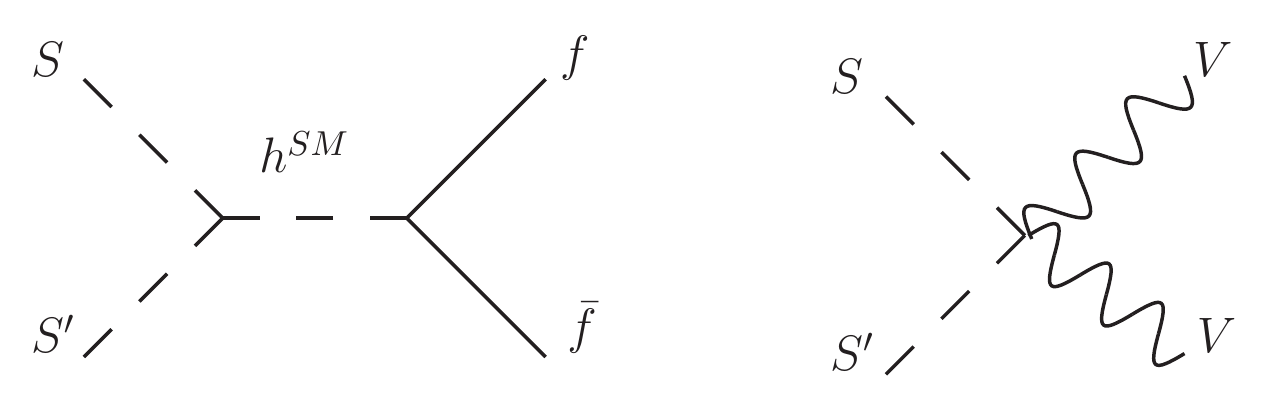}
\caption{If all inert particles are very close in mass, all the above coannihilation channels contribute to the relic density value, where $S,S' = H_i, A_i, H^\pm_i$.}
\label{annihilation-5}
\end{figure}

If all inert particles are very close in mass then all channels shown in Fig.(\ref{annihilation-5}) contribute to the final DM relic density.

\subsection{Coannihilation scenarios}\label{annihilation-scenarios}

We introduce the following parameters for the mass splitting between the DM candidate $H_1$ and other inert particles:
\begin{itemize}
\item 
between $H_1$ and the other CP-even state $H_2$
\be 
\Delta = m_{H_2} - m_{H_1}
\ee

\item 
between $H_1$ and the pseudoscalar/charged state from the (same) lighter generation
\be 
\delta_A = m_{A_1} -m_{H_1} , \quad \delta_C = m_{H^\pm_1} - m_{H_1},
\ee

\item 
between $H_1$ and pseudoscalar/charged state from the (other) heavier generation
\be 
\delta'_A = m_{A_2} -m_{H_1} , \quad \delta'_C = m_{H^\pm_2} - m_{H_1}.
\ee
\end{itemize}

The following scenarios are relevant for DM relic density studies:
\begin{itemize}
\item[A)] \textbf{Large $\Delta,\delta_A, \delta_C \Rightarrow $ large $\delta'_A, \delta'_C$}
\\The DM particle is significantly lighter than other inert particles. This case is similar to the standard Higgs-portal approach, especially if $m_{H_1}$ is very light. One should, however, take into account the importance of annihilation channels with virtual gauge bosons even in the $m_{\rm DM} < m_W$ region.

\item[B)] \textbf{Small $\Delta$ and large $\delta_A, \delta_C \Rightarrow $ large $\delta'_A, \delta'_C$}
\\There is a small difference between the masses of $H_1$ and $H_2$, thus the coannihilation between those two particles occurs through $H_1 H_2 \to h \to \bar{f}f$ while there is no coannihilation between scalar and pseudo-scalar particles.

\item[C)] \textbf{Small $\delta_A$ and large $\Delta,\delta_C \Rightarrow $ large $\delta'_A, \delta'_C$}
\\There is a small difference between the masses of $H_1$ and $A_1$ while the CP-even particles mass splitting is large. The situation in this case is similar to the I(1+1)HDM with small CP-even/CP-odd mass splitting.

\item[D)] \textbf{Small $\delta_A,\Delta$, large $\delta_C\Rightarrow $ large $ \delta'_C$, while $\delta'_A$ can be small}
\\There is a small difference between the masses of $H_1$, $A_1$, $H_2$ and possibly $A_2$ (depending on the size of  $\delta_A$ and $\Delta$), which can lead to coannihilation between all neutral inert particles. 

The above scenarios are the only phenomenologically relevant cases in the $m_{\rm DM} < m_{h}/2$ region. This is due to the LEP limits on the charged scalar excluding $m_{H_i}^\pm < 70-90$ GeV, which means that it cannot be close in mass with the DM candidate in this mass region. 

However, for $m_{\rm DM} > m_W$, coannihilation with $H^\pm_i$ is  also allowed leading to the two following cases, which are analogous scenarios to C and D  with $\delta_A (') \leftrightarrow \delta_C (')$:

\item[E)] \textbf{Small $\delta_C$ and large $\Delta,\delta_A \Rightarrow $ large $ \delta'_A,\delta'_C$}
\item[F)] \textbf{Small $\delta_C,\Delta$, large $\delta_A\Rightarrow $ large $ \delta'_A$, while $\delta'_C$ can be small}

These two scenarios can be realized in the $m_{\rm DM}>m_W$ region which is dominated by very effective annihilation into gauge bosons. Usually one needs strong cancellation effects to obtain proper relic density.

A final possible coannihilation scenario is the following.

\item[G)] \textbf{Small $\Delta,\delta_A,\delta_C \Rightarrow $ possible small $ \delta'_C, \delta'_A$}
\\All inert particles have similar masses, which resembles the situation in the heavy mass region in the I(1+1)HDM.
\end{itemize}

In principle, a scalar DM candidate with acceptable relic density abundance can have a mass range from a few GeV to a few TeV. A light DM candidate usually requires a relatively large DM-Higgs coupling. As the mass grows, the annihilation into Higgs bosons becomes more effective, especially around the Higgs resonance ($m_{\rm DM} \approx m_{h}/2$) and a smaller DM-Higgs coupling is required. 
For $m_{\rm DM} > m_W$ annihilation into gauge bosons is very effective which usually leads to relic density values well below the observed limit. The cancellation effects between diagrams in Fig.(\ref{annihilation-2}) and coannihilation effects can help restore the relic density value. 

In section \ref{constrained-3IDM} we will numerically investigate all possible scenarios for DM (co)annihilation in the I(2+1)HDM in different regions of DM mass. Here, we briefly comment on the gross feature of each benchmark configuration.

For a start, note that, in the I(2+1)HDM, DM-nucleon scattering is through the exchange of the Higgs particle in the $t$-channel. This cross section depends on the Higgs-DM coupling, $g_{H_1H_1h}$, and, as discussed, in many cases in the Higgs-portal models the coupling needed for the proper relic density is too large to reconcile with results from direct detection experiments. As it will be shown in section \ref{constrained-3IDM}, it is often the case in scenario A.

Furthermore, note that the coannihilation effects may help in satisfying these constraints. In scenarios B--G an acceptable value of relic density is obtained through coannihilation processes, while the scattering still depends only on the $t$-channel Higgs-exchange. DM-Higgs coupling in such cases may be smaller than in case A, and so the scattering cross section may lie below the current experimental limits.

Finally, note that the direct detection experiments set strong limits on the scattering through $Z$-exchange in regions where scalar and pseudoscalar states are (nearly) degenerate. A non-zero mass splitting between the lightest scalar and the lightest pseudoscalar particle needs to be larger than the kinetic energy of DM in our galactic halo, so that scattering through $Z$-exchange is forbidden kinematically. This sets the lower limit on $\delta_A$ to be a few 100 keV \cite{Akerib:2005kh}. In principle, this limit has important consequences for NHDMs, where degenerated states appear in a natural way. In the I(2+1)HDM though, the degeneracy between $H_i$ and $A_i$ is lifted by non-zero $\lambda_{2}$ and $\lambda_3$.

\section{The simplified I(2+1)HDM }\label{constrained-3IDM}

\subsection{Simplified couplings in the I(2+1)HDM}
We study the simplified case of the I(2+1)HDM where all the parameters related to the first inert doublet are $k$ times the parameters related to the second doublet:
\be 
\mu^2_1 = k \mu^2_2, \quad \lambda_3 = k \lambda_2, \quad \lambda_{31} = k \lambda_{23}, \quad \lambda_{31}' = k \lambda_{23}', 
\ee
resulting in
\be 
\Lambda_{\phi_1} = k \Lambda_{\phi_2}, \quad \Lambda'_{\phi_1} = k\Lambda'_{\phi_2}, \quad \Lambda''_{\phi_1} = k \Lambda''_{\phi_2}.
\ee
However, we assume no specific relation among the other parameters of the potential.

The masses of the CP-even neutral states in this case are simplified to
\bea
&& m^2_{H_1} = (-\mu^2_2+\Lambda_{\phi_2})(k\cos^2\theta_h + \sin^2\theta_h) -2\mu^2_{12}\sin\theta_h \cos\theta_h \nonumber\\
&& m^2_{H_2} = (-\mu^2_2+\Lambda_{\phi_2})(k\sin^2\theta_h + \cos^2\theta_h) +2\mu^2_{12}\sin\theta_h \cos\theta_h \nonumber
\eea
with the mixing angle between the CP-even states given by
\be 
\tan 2\theta_h = \frac{-2\mu^2_{12}}{(k-1)(-\mu^2_2 + \Lambda_{\phi_2})}, 
\label{cp-even-mixing}
\ee
$m^2_{H^\pm_2}, m^2_{H^\pm_1}$ and $m^2_{A_2}, m^2_{A_1}$ have similar values with $\Lambda_{\phi_2}$ and $\theta_h$ replaced by $\Lambda'_{\phi_2}$, $\Lambda''_{\phi_2}$ and $\theta_c$, $\theta_a$, respectively.

Note that the positivity of mass eigenstates puts the following limits on the acceptable values of $k$:
\bea 
& (\frac{2\mu^2_{12}}{-\mu^2_2 + \Lambda_{\phi_2}}- \tan\theta_h )\tan\theta_h \quad < \quad k \quad < \quad(\frac{2\mu^2_{12}}{-\mu^2_2 + \Lambda_{\phi_2}}+ \cot\theta_h ) \cot\theta_h, &\\[2mm]
& (\frac{2\mu^2_{12}}{-\mu^2_2 + \Lambda'_{\phi_2}}- \tan\theta_c )\tan\theta_c \quad < \quad k \quad < \quad(\frac{2\mu^2_{12}}{-\mu^2_2 + \Lambda'_{\phi_2}}+ \cot\theta_c ) \cot\theta_c, &\nonumber\\[2mm]
& (\frac{2\mu^2_{12}}{-\mu^2_2 + \Lambda''_{\phi_2}}- \tan\theta_a )\tan\theta_a \quad < \quad k \quad < \quad(\frac{2\mu^2_{12}}{-\mu^2_2 + \Lambda''_{\phi_2}}+ \cot\theta_a ) \cot\theta_a. &\nonumber
\eea

We will study several cases in the simplified I(2+1)DM which are listed in Tab.(\ref{cases-table}). First, cases of $k=0$ (section \ref{section-IDM}) and $k=1$ with $\mu_{12}^2=0$ (section \ref{section-keq1NM}) are discussed briefly for completeness, but they do not provide any solution to the problems of the Higgs-portal DM scenario. We then study the case of $k=1$ with $\mu^2_{12} \neq 0$ in detail since it represents all features of this model clearly. Our numerical DM studies are done mostly for the selected benchmark points which exhibit typical behaviour of a particular scenario. In each case we present the resulting relic density plots and in section \ref{summary-plots} the different cases are compared to each other.
The other cases from Tab.(\ref{cases-table}) are discussed briefly since they do not present any new features of the model.

\begin{table} [ht!]
\begin{footnotesize}
\begin{center}
\begin{tabular}{|m{2cm} | m{3cm}| m{6cm}| m{1cm}|} \hline
{k}                & \mbox{mixing}   &  \mbox{parameters} & \mbox{section}   \\[3mm] \hline  \hline
$k=0 $ & $NA$         & $I(1+1)HDM$  & \ref{section-IDM}      
\\[3mm] \hline
$k=1$  & $\mu^2_{12}=0$ & $m^2_{H_1},m^2_{A_1},m^2_{H^\pm_1},g_{H_1H_1h}$  & \ref{section-keq1NM}     \\[3mm] \hline
$k=1$  & $\mu^2_{12}\neq 0$          & $m^2_{H_1},m^2_{A_1},m^2_{H^\pm_1},g_{H_1H_1h},\mu^2_{12}$  & \ref{section-keq1M}     \\[3mm] \hline  
$k\neq 1$  & $\mu^2_{12}=0$ & $m^2_{H_1},m^2_{A_1},m^2_{H^\pm_1},g_{H_1H_1h}, k $  & \ref{section-kgr1NM}  
\\[3mm]\hline
$k\neq 1$  & $\mu^2_{12} \neq 0$  & $m^2_{H_1},m^2_{A_1},m^2_{H^\pm_1},g_{H_1H_1h},\mu^2_{12}, k$ & \ref{section-kgr1M}  
\\[3mm] \hline   
\end{tabular}
\end{center}
\end{footnotesize}
\caption{The cases with different values of $k$ and mixing between the inert doublets studied here alongside the paper section where they are dealt with.}
\label{cases-table}
\end{table}

\subsection{The $k=0$ case}\label{section-IDM}
With $k=0$ the model is reduced to the two doublet case, the I(1+1)HDM, which we briefly review here.
The $Z_2$-symmetric I(1+1)HDM potential of two doublets, one active and one inert, is commonly written as:
\bea
V^{I(1+1)HDM} &=& - \mu^2_{1} (\phi_1^\dagger \phi_1) -\mu^2_2 (\phi_2^\dagger \phi_2) + \lambda_{1} (\phi_1^\dagger \phi_1)^2+ \lambda_{2} (\phi_2^\dagger \phi_2)^2 \\
&& + \lambda_{3}  (\phi_1^\dagger \phi_1)(\phi_2^\dagger \phi_2) + \lambda_{4} (\phi_1^\dagger \phi_2)(\phi_2^\dagger \phi_1) +  \lambda_{5}(\phi_1^\dagger\phi_2)^2 +  h.c. \nonumber
\eea
with all real parameters and $g_{Z_2}=\mbox{diag}(1,-1)$. This symmetry is respected by the vacuum alignment $(v,0)$ and the neutral fields from the inert doublet, $\phi_2$, are viable DM candidates.
In the $\lambda_5 <0$ and $\lambda_4 + \lambda_5 <0$ region, the DM candidate, i.e., the lightest $Z_2$-odd particle, would be the CP-even neutral particle $H$.



In this model there are three distinctive regions of $m_{H}$ where one can expect to obtain proper relic density \cite{Barbieri:2006dq,LopezHonorez:2006gr,Cao:2007rm,Dolle:2009fn,Dolle:2009ft,Arina:2009um, Tytgat:2007cv, Honorez:2010re,LopezHonorez:2010tb,Sokolowska:2011aa,Sokolowska:2011sb}.
\begin{enumerate}
\item[{(a)}] A light DM candidate with mass $\lesssim 10 \textrm{ GeV}$, where DM annihilates mostly into $b\bar{b}$ through the Higgs exchange.
\item[{(b)}] A medium DM mass of $40-150 \textrm{ GeV}$ with or without coannihilation with the CP-odd state $A$.
\item[{(c)}] A heavy state DM  with mass $\gtrsim 550 \textrm{ GeV}$, where all particles' masses are almost degenerate and relic density is driven by coannihilation processes combined with annihilation into gauge bosons. 
\end{enumerate} 

The I(1+1)HDM parameter space is strongly constrained by the recent LHC and direct detection results which lead to exclusion of  the low DM mass region for $m_{\rm DM} \lesssim 55$ GeV \cite{Krawczyk:2013jta, Goudelis:2013uca,Arhrib:2013ela}. This is due to the incompatibility between the relic density limits, which require the Higgs-DM coupling to be relatively big to ensure the efficient-enough DM annihilation, and the LHC $\mu_{\gamma\gamma}$ and $Br(h\to inv.)$ constraints, which prefer much smaller values of such a coupling.
The
medium mass region for $m_{\rm DM} < m_W$, is instead in agreement with the current experimental results, however, an enhancement in the diphoton Higgs decay channel is disfavored \cite{Krawczyk:2013jta}. Finally, LHC and direct detection limits do not provide significant constraints on the heavy mass region, however, if a significant enhancement in $\mu_{\gamma\gamma}$ is confirmed, then it is not possible to reconcile it with relic density limits.

\subsection{The $k=$1 case with vanishing mixing}\label{section-keq1NM}

In this case the two inert doublets are completely degenerate:
\be 
\mu^2_1 = \mu^2_2, \quad \lambda_2 = \lambda_3, \quad \lambda_{23} = \lambda_{31}, \quad \lambda_{23}' = \lambda_{31}'. 
\ee
Furthermore, the $\mu^2_{12} =0$ leaves only 4 base parameters,
\be 
\mu^2_2, \lambda_2, \lambda_{23}, \lambda'_{23},  
\ee
and the 5 parameters describing the self-interactions of inert particles, i.e., $\lambda_2, \lambda_{11},\lambda_{22},\lambda_{12}, \lambda'_{12}$, which are not relevant for the relic density analysis.

Note that the potential becomes $Z_2 \times Z_2$ symmetric after imposing these equalities, as

\bea
V &=& - \mu^2_{2} (\phi_1^\dagger \phi_1 + \phi_2^\dagger \phi_2) - \mu^2_3(\phi_3^\dagger \phi_3) \\
&&+ \lambda_{22} \biggl((\phi_1^\dagger \phi_1)^2+  (\phi_2^\dagger \phi_2)^2 \biggr)  + \lambda_{33} (\phi_3^\dagger \phi_3)^2 \nonumber\\
&& + \lambda_{23} \biggl( (\phi_1^\dagger \phi_1)(\phi_2^\dagger \phi_2)
 +   (\phi_2^\dagger \phi_2)(\phi_3^\dagger \phi_3)\biggr) + \lambda_{31} (\phi_3^\dagger \phi_3)(\phi_1^\dagger \phi_1) \nonumber\\
&& + \lambda'_{23} \biggl((\phi_1^\dagger \phi_2)(\phi_2^\dagger \phi_1) 
 +  (\phi_2^\dagger \phi_3)(\phi_3^\dagger \phi_2)\biggr) + \lambda'_{31} (\phi_3^\dagger \phi_1)(\phi_1^\dagger \phi_3)  \nonumber\\
&& +  \lambda_{1}(\phi_1^\dagger\phi_2)^2 + \lambda_2\biggl((\phi_2^\dagger\phi_3)^2 + (\phi_3^\dagger\phi_1)^2\biggr)  + h.c. \nonumber
\eea

The inert particle mass spectrum in this case has the following form:
\bea
&& m^2_{H_2} = m^2_{H_1} = -\mu^2_2 +\Lambda_{\phi_2},\\
&& m^2_{H^\pm_2} =  m^2_{H^\pm_1} = -\mu^2_2 +\Lambda'_{\phi_2}, \nonumber \\
&& m^2_{A_2} =  m^2_{A_1} = -\mu^2_2 +\Lambda''_{\phi_2}. \nonumber
\eea

The parameters of the model in terms of the physical parameters are:
\bea
&& \lambda'_{23} = \frac{1}{v^2} (m_{H_1}^2 +m_{A_1}^2- 2m_{H^\pm_1}^2),  \\
&& \lambda_2 = \frac{1}{2v^2}(m_{H_1}^2-m_{A_1}^2), \nonumber\\
&& \lambda_{23} = \frac{1}{v^2} (2m_{H^\pm_1}^2 - 2m_{H_1}^2) + g_{H_1 H_1 h},  \nonumber\\
&& \mu^2_2 =\frac{v^2}{2}g_{H_1 H_1 h}  - m^2_{H_1}. \nonumber
\eea

The relevant Feynman rules are:
\bea
&& H^+_2 H^-_2, H^+_1 H^-_1 \longrightarrow h \qquad \qquad \lambda_{23} v \\
&& H_1 H_1, H_2 H_2 \longrightarrow h \qquad \qquad (\lambda_{23}+ \lambda'_{23} +2\lambda_{2}) \frac{v}{2} \nonumber\\
&& A_1 A_1, A_2 A_2 \longrightarrow h \qquad \qquad (\lambda_{23} + \lambda'_{23} -2\lambda_{2} ) \frac{v}{2} \nonumber\\
&& H^+_2 H^-_2, H^+_1 H^-_1 \longrightarrow \gamma \qquad \qquad \frac{i}{2}(g\sin\theta_W + g'\cos\theta_W)(K+K')^\mu \nonumber\\
&& H^+_2 H^-_2, H^+_1 H^-_1 \longrightarrow Z \qquad \qquad \frac{i}{2}(g\cos\theta_W - g'\sin\theta_W)(K+K')^\mu \nonumber\\
&& H^\pm_1 H_1, H^\pm_2 H_2 \longrightarrow W^\pm \qquad \qquad \frac{ig}{2}(K+K')^\mu \nonumber\\
&& H^\pm_1 A_1, H^\pm_2 A_2  \longrightarrow W^\pm \qquad \qquad  \frac{g}{2} (K+K')^\mu \nonumber\\
&& H_1 A_1, H_2 A_2 \longrightarrow Z \qquad \qquad   \frac{1}{2}(g\cos\theta_W + g'\sin\theta_W)(K+K')^\mu \nonumber
\eea

In first approximation, i.e., by neglecting the self-interactions of the inert doublets (for example, $H_1 H_1 \leftrightarrow H_2 H_2$), one can treat this case as a doubled I(1+1)HDM, with two DM candidates, $H_1$ and $H_2$. They have degenerated masses and identical interactions, as noted above, and they  contribute equally to the DM relic density $\Omega_{DM}h^2$:
\begin{equation}
\Omega_{DM}h^2 = \Omega_{H_1}h^2 + \Omega_{H_2}h^2 = 2 \Omega_{H}h^2,
\end{equation}
where $\Omega_{H}h^2$ is the relic density of a single DM candidate from the I(1+1)HDM.

To fulfil the Planck limit, $\Omega_{H}h^2$ needs to lie between $0.0559$ and $0.064$, meaning that the DM annihilation should be more effective than in the I(1+1)HDM. This requires even bigger values of the DM-Higgs coupling, which, for $m_{\rm DM} < m_h/2$, would lead to even larger values of $Br(h\to inv.)$, making it even harder to satisfy relic density and LHC bounds simultaneously, unless it is in the Higgs-resonance region.

\subsection{The $k=1$ case with non-vanishing mixing}\label{section-keq1M}
Similar to the previous case, the two inert doublets are perfect copies of each other, with equal parameters.
The non-zero CP-even mixing angle from Eq.(\ref{cp-even-mixing}) is
\be 
\tan 2\theta_h = \pm \infty \quad \rightarrow \quad \theta_h = \pm \pi/4.
\ee
The CP-even mass spectrum is therefore of the following form:
\bea 
&& m_{H_1}^2 = (- \mu_2^2 + \Lambda_{\phi_2}) -2 \mu_{12}^2 \sin\theta_h \cos\theta_h,  \\
&& m_{H_2}^2 = (- \mu_2^2 + \Lambda_{\phi_2}) +2 \mu_{12}^2 \sin\theta_h \cos\theta_h.  \nonumber
\eea
Assuming $\mu^2_{12} >0$ and $\theta_h$ belonging to the $1^{\rm st}$ quadrant, one has
\be  
\sin\theta_h = \cos\theta_h = \frac{\sqrt{2}}{2} \quad \rightarrow\quad m^2_{H_1} < m^2_{H_2} 
\ee
which indeed makes $H_1$ the lightest among the inert particles and therefore our DM candidate.

Further assuming that all $\theta_i$'s are in the $1^{\rm st}$ quadrant, the mass spectrum has the following form:
\bea 
&& m_{H_1}^2 = - \mu_2^2 + \Lambda_{\phi_2} - \mu_{12}^2, \qquad m_{H_2}^2 =  m_{H_1}^2 + 2\mu_{12}^2,\\
&& m_{H^\pm_1}^2 = - \mu_2^2 + \Lambda'_{\phi_2} - \mu_{12}^2, \qquad m_{H^\pm_2}^2 = m_{H^\pm_1}^2 + 2\mu_{12}^2, \nonumber\\
&& m_{A_1}^2 = - \mu_2^2 + \Lambda''_{\phi_2} - \mu_{12}^2 , \qquad m_{A_2}^2  = m_{A_1}^2 + 2\mu_{12}^2. \nonumber
\eea

The base parameters can then be expressed in terms of 
\be 
m_{H_1, A_1, H^\pm_1}, \quad \mu_{12}^2 \; (\textrm{or equivalently } m_{H_2} \textrm{ or } \Delta), \quad g_{H_1 H_1 h}.
\ee

Finally, the following equations relate different parameters:
\bea
&& \lambda_{23}' = \frac{1}{v^2} (m_{H_1}^2 +m_{A_1}^2 -2m_{H^\pm_1}^2), \\
&& \lambda_2 = \frac{1}{2v^2} (m_{H_1}^2 - m_{A_1}^2), \nonumber\\
&& \lambda_{23} = g_{H_1 H_1 h} - \frac{2}{v^2}(m_{H_1}^2 - m_{H^\pm_1}^2), \nonumber\\ 
&& \mu^2_2 =\frac{v^2}{2} g_{H_1 H_1 h} - m_{H_1}^2 - \mu^2_{12}, \nonumber\\
&& \mu_{12}^2 = \frac{1}{2}(m_{H_2}^2 - m_{H_1}^2) = \frac{1}{2} \left(\Delta^2+2 m_{H_1} \Delta\right),\nonumber
\eea
where $g_{H_1 H_1 h}\frac{v}{2}$ is the coefficient of the $H_1 H_1 h$ term in the potential.

It is interesting to note that the equations for $\lambda'_{23}$ and $\lambda_2$ are identical to the corresponding relation for $\lambda_4$ and $\lambda_5$ in the I(1+1)HDM case and that the phenomenology of the model depends on masses of inert particles and one coupling only.

The relevant Feynman rules are:
\bea
&& H^+_2 H^-_2, H^+_1 H^-_1 \longrightarrow h \qquad \qquad \lambda_{23} v \nonumber\\
&& H_1 H_1, H_2 H_2 \longrightarrow h \qquad \qquad (\lambda_{23}+ \lambda'_{23} +2\lambda_{2}) \frac{v}{2} \nonumber\\
&& A_1 A_1, A_2 A_2 \longrightarrow h \qquad \qquad (\lambda_{23} + \lambda'_{23} -2\lambda_{2} ) \frac{v}{2} \nonumber\\
&& H^+_2 H^-_2, H^+_1 H^-_1 \longrightarrow \gamma \qquad \qquad \frac{i}{2}(g\sin\theta_W + g'\cos\theta_W)(K+K')^\mu \nonumber\\
&& H^+_2 H^-_2, H^+_1 H^-_1 \longrightarrow Z \qquad \qquad \frac{i}{2}(g\cos\theta_W - g'\sin\theta_W)(K+K')^\mu \nonumber\\
&& H^\pm_1 H_1, H^\pm_2 H_2 \longrightarrow W^\pm \qquad \qquad \frac{ig}{2}\cos(\theta_h -\theta_c)(K+K')^\mu \nonumber\\
&& H^\pm_2 H_1, H^\pm_1 H_2 \longrightarrow W^\pm \qquad \qquad \frac{ig}{2}\sin(\theta_h -\theta_c)(K+K')^\mu \nonumber\\
&& H^\pm_1 A_1, H^\pm_2 A_2  \longrightarrow W^\pm \qquad \qquad  \frac{g}{2} \cos(\theta_a -\theta_c)(K+K')^\mu \nonumber\\
&& H^\pm_2 A_1, H^\pm_1 A_2  \longrightarrow W^\pm \qquad \qquad  \frac{g}{2} \sin(\theta_a -\theta_c)(K+K')^\mu \nonumber\\
&& H_1 A_1, H_2 A_2 \longrightarrow Z \qquad \qquad   \frac{1}{2}(g\cos\theta_W + g'\sin\theta_W)\cos(\theta_h -\theta_a)(K+K')^\mu \nonumber\\
&& H_2 A_1, H_1 A_2 \longrightarrow Z \qquad \qquad   \frac{1}{2}(g\cos\theta_W + g'\sin\theta_W)\sin(\theta_h -\theta_a)(K+K')^\mu \nonumber
\eea


In the following subsections, we study the $k=1$ with $\mu^2_{12}\neq 0$ case in detail for the scenarios proposed in section \ref{annihilation-scenarios}. Firstly, scenarios with at least one open invisible Higgs decay channel, i.e. $m_{H_1} < m_h/2$ are discussed. The lower limit for the DM mass is taken to be $m_{H_1} \approx m_Z/2 \approx 45$ GeV. Other invisible channels, $h\to A_1 A_1, A_2 A_2, H_2 H_2$ may be open, which is the case in scenarios B-D, where important coannihilation effects are present.

In the second part of this section, scenarios with DM mass from the $m_W > m_{H_1} > m_h/2$ range are discussed where the effects of DM annihilation into gauge bosons play an important role and lead to a rather different phenomenology.


\subsubsection{Open invisible channels ($m_{\rm DM} < m_h/2$)}

\begin{figure}[h!]
\centering
\includegraphics[width=0.8\textwidth]{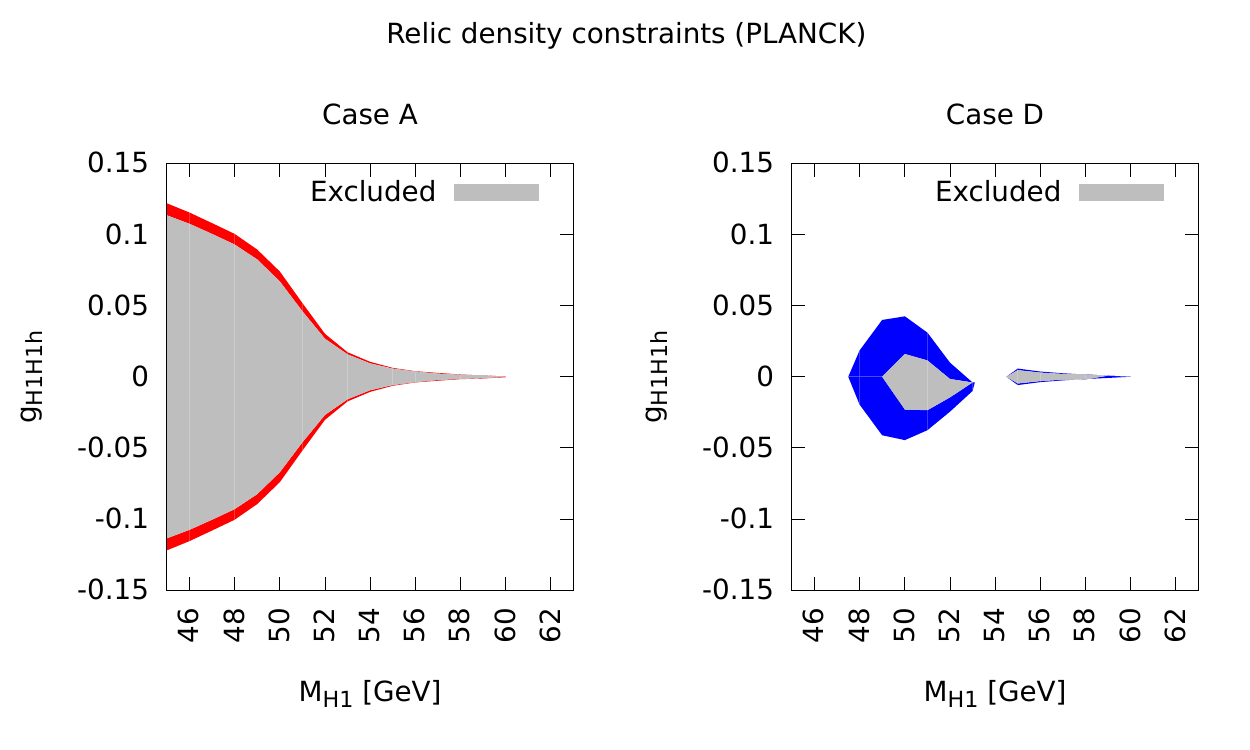} 
\caption{Relic density constraints on the mass of the DM candidate and its coupling to SM Higgs boson, with the white and gray regions representing too little and too much relic abundance, respectively. The red band is where sufficient amount of relic abundance is produced in case A (and identically in case B$_8$), and the blue region shows the accepted window in case D. In case C the relic density is always below the observed value.}
\label{Mass-coupling}
\end{figure}

\begin{itemize}
\item
\textbf{Case A with $\Delta = 50$ GeV}\\
We choose the following values of masses as an example for our numerical studies:  
\begin{equation}
m_{A_1} = 115 \mbox{ GeV}, \qquad m_{H^\pm_1} = 115 \mbox{ GeV}, \qquad m_{H_2} = m_{H_1} + 50 \mbox{ GeV}.
\end{equation}
Mass splittings $\delta_A, \delta_C$ are of the order of 50 GeV, with $H_1$ being much lighter than other inert particles. The resulting relic density is plotted versus the DM-Higgs coupling in Fig.(\ref{plotA}), which shows that for DM masses below $m_h/2$ a small DM-Higgs coupling, $g_{H_1 H_1 h}$, usually leads to very large relic densities, since small $|g_{H_1 H_1 h}|$ corresponds to a slow annihilation rate $H_1 H_1 \rightarrow h$. 
The smaller $m_{H_1}$ is, the larger $g_{H_1 H_1 h}$ coupling is needed in order to produce a relic density in agreement with observation. In the Higgs-resonance region, where $m_{H_1} \approx m_h/2$, viable DM-Higgs coupling has drastically smaller values.

The scenario in case A for $m_{\rm DM}<m_h/2$ can be considered as the purest Higgs-portal-type case provided by the I(2+1)HDM, with all its disadvantages; large DM-Higgs coupling values, which are needed for efficient annihilation, are in tension with direct detection limits and invisible Higgs decay constraints from the LHC. 
Small DM-Higgs couplings lead to large relic density, however, this DM abundance can be reduced, provided other DM annihilation channels are open. For large enough $m_{H_1}$, DM can annihilate through gauge bosons. However, the existence of these channels is mass- and case-dependent (see also Fig.(\ref{Mass-coupling})). 

Note that, if the mass splittings $\delta_{A,C},\delta'_{A,C}$ are large enough to forbid any coannihilation effects, the lightest $Z_2$-odd particle is effectively decoupled from the $Z_2$-odd sector and the exact values of masses do not influence the DM phenomenology. In that sense DM studies, i.e., relic density measurements and direct detection experiments, do not put any additional constraints on heavier $Z_2$-odd particles and input from particle physics is needed.

\begin{figure}[ht!]
\centering
\subfloat[]{\label{plotA}\includegraphics[width=0.45\textwidth]{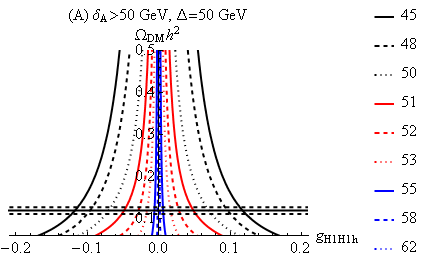}} \quad
\subfloat[]{\label{A-Masses}\includegraphics[width=0.4\textwidth]{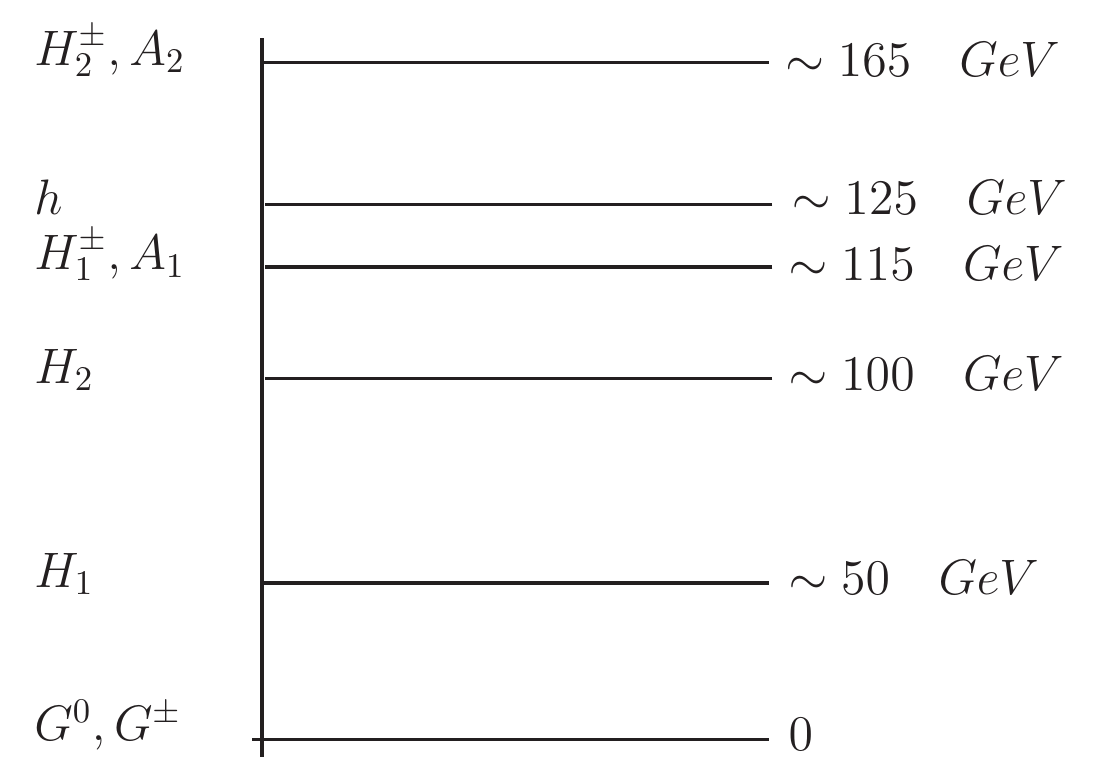}} \\
\caption{Case A. (a) Relic density vs. DM-Higgs coupling is shown. Here $m_{H_1}$ changes from $45$ to $62$ GeV. Horizontal lines denote the Planck value $\Omega_{DM} h^2 = 0.1199 \pm 3 \sigma$, the region above is excluded. (b) Schematic mass spectrum for which a mid value of $m_{H_1}$ has been chosen and masses are roughly approximated.} 
\label{scan11}
\end{figure}

\item
\textbf{Case B with $\Delta = 8$ GeV (case B$_8$)}\\

In this scenario, with $\Delta$ relatively small, one expects the $H_2H_1 \rightarrow h$ coannihilation effects to show in the relic density plots. Note that the result is sensitive to the value of $\Delta$. Here we show the relic density plot for the $\Delta = 8$ GeV case (referred to as case B$_8$) and comment on the $\Delta = 1$ GeV case (referred to as case B$_1$).

One should note first that for the discussed $k=1$ case there is no tree-level $H_2 H_1 h$ coupling and so $H_1 H_2 \to h \to f \bar{f}$ diagrams do not exist. Therefore, in comparison with case A the annihilation of DM is not affected, which is shown in the the first considered setup, $\Delta = 8$ GeV, presented in Fig.(\ref{plotB8}). Note that  Fig.(\ref{plotB8}) is almost identical to case A in Fig.(\ref{plotA}). The reason being that the $\Delta$ mass splitting is large enough and so the $H_2 H_2 \to h \to f \bar{f}$ diagrams do not interfere with the thermal evolution of DM relic density. 

\textbf{Case B with $\Delta = 1$ GeV (case B$_1$)}\\
With smaller mass splitting, $\Delta \approx 1$ GeV, the relic density evolution could be affected.
In this case the second CP-even particle $H_2$ acts almost like the second DM candidate discussed in the $k=1$ no-mixing scenario in section \ref{section-keq1NM}. The obtained relic density is larger than in case A and larger couplings are needed to fulfil the Planck bounds. This is even harder to reconcile with limits from $Br(h\to inv.)$, especially since now there are two invisible channels open. Therefore, we conclude that scenario B cannot provide any solution to the problems of Higgs-portal DM with $m_{\rm DM}<m_h/2$.

\begin{figure}[ht!]
\centering
\subfloat[]{\label{plotB8}\includegraphics[width=0.45\textwidth]{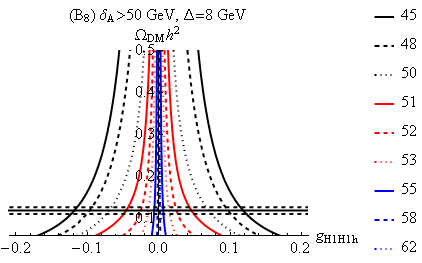}} \quad
\subfloat[]{\label{B-Masses8}\includegraphics[width=0.4\textwidth]{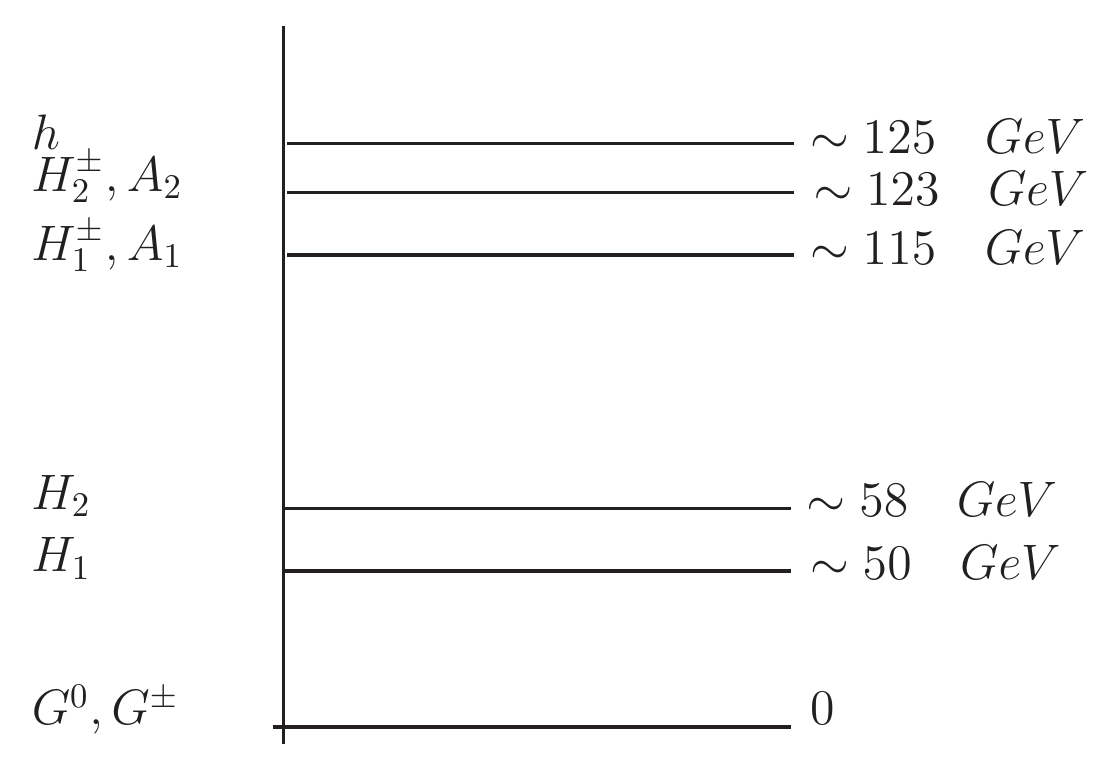}} \\
\caption{Case B$_8$ with $\Delta = 8$ GeV. (a) Relic density vs. DM-Higgs coupling is shown. Here, $m_{H_1}$ changes from $45$ to $62$ GeV. Horizontal lines denote the Planck value $\Omega_{DM} h^2 = 0.1199 \pm 3 \sigma$, the region above is excluded. (b) Schematic mass spectrum for which a mid value of $m_{H_1}$ has been chosen and masses are roughly approximated.}
\end{figure} 


\item 
\textbf{Case C with $\delta_A = 8$ GeV and $\Delta = 50$ GeV}\\

In this case $H_1$ and $A_1$ are very close in mass, while other inert particles are heavy in comparison: 
\begin{equation}
m_{A_1} = m_{H_1} + 8 \mbox{ GeV}, \quad m_{H^\pm_1} = m_{H_1} + 50 \mbox{ GeV}, \quad m_{H_2} = m_{H_1} + 50 \mbox{ GeV}.
\end{equation}
As a result, there is coannihilation between $H_1$ and $A_1$ whose effects are visible in Fig.(\ref{plotC}) compared to the previous cases.
 
For this setup coannihilation effects lead to an enhanced cross section. Coannihilation becomes so effective that, even for small couplings, relic density does not reach the observed value. In fact, coannihilation processes are so strong that for every value of DM mass (below $m_{h}/2$) relic density is below current Planck/WMAP limits. This situation does not result in the exclusion of this parameter space but rather corresponds to a subdominant DM candidate.

Note that the $\delta_A$ chosen here is the boundary value of mass splitting between scalar and pseudoscalar in the I(1+1)HDM, obtained by translation of the null-searches for charginos and neutralinos at LEP-II \cite{Lundstrom:2008ai}. Increasing this value to $\delta_A = 10$~GeV and thus reducing the strength of coannihilation - while still keeping it possible - allows for $\Omega_{DM}h^2$ within Planck limits for relatively small values of DM-Higgs coupling. 
    
\begin{figure}[ht!]
\centering    
\subfloat[]{\label{plotC}\includegraphics[width=0.45\textwidth]{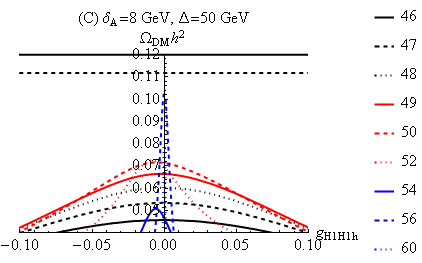}} \quad
\subfloat[]{\label{C-Masses}\includegraphics[width=0.4\textwidth]{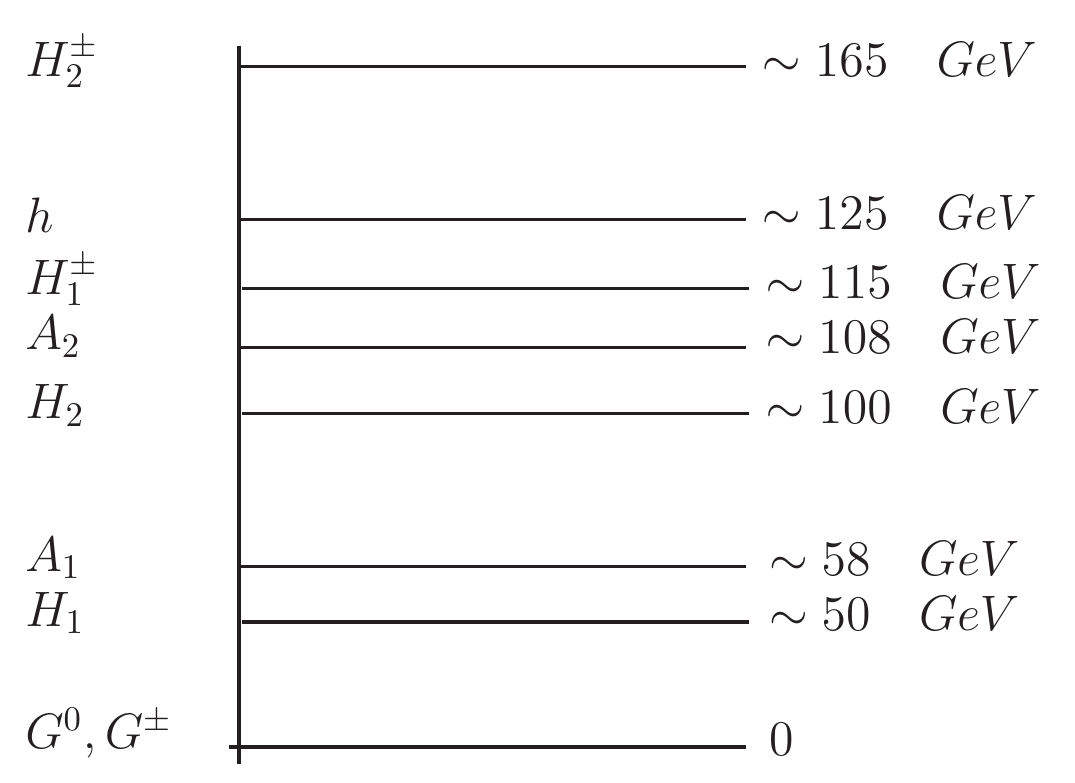}} \\
\caption{Case C. (a) Relic density vs. DM-Higgs coupling $g_{H_1H_1h}$ is shown. Here, $m_{H_1}$ changes from $45$ to $62$ GeV. Horizontal lines denote the Planck value $\Omega_{DM} h^2 = 0.1199 \pm 3 \sigma$, the region above is excluded. (b) The schematic mass spectrum for which a mid value of $m_{H_1}$ has been chosen and masses are roughly approximated.} 
\label{scan12}
\end{figure}

\item 
\textbf{Case D with $\delta_A = 7$ GeV and $\Delta = 1$ GeV}\\

In this case the masses of all neutral inert particles $H_{2,1}, A_{2,1}$ are relatively close. Two important coannihilation effects taking place here are the following. 
Firstly, 
the $H_1 A_1 \rightarrow Z$ coannihilation, which leads to a decrease in the relic density (similar to most SUSY models). As discussed in case C above, usually it leads to $\Omega_{DM}h^2$ below the observed value.
Secondly,
the $H_2 H_2 \rightarrow h$ (co)annihilation effect, which leads to an increase in the relic density (similar to the case in UED theories) by affecting the DM production rate, as presented in case B$_1$ with small $\Delta$. 
%
These two effects combined will allow for small $g_{H_1H_1h}$ values and result in sufficient relic density, which is desirable since smaller $g_{H_1H_1h}$ leads to less stringent bounds from direct detection experiments.

\begin{figure}[ht!]
\centering
\subfloat[]{\label{plotD}\includegraphics[width=0.45\textwidth]{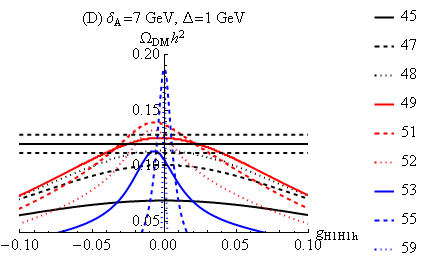}} \quad
\subfloat[]{\label{D-Masses}\includegraphics[width=0.4\textwidth]{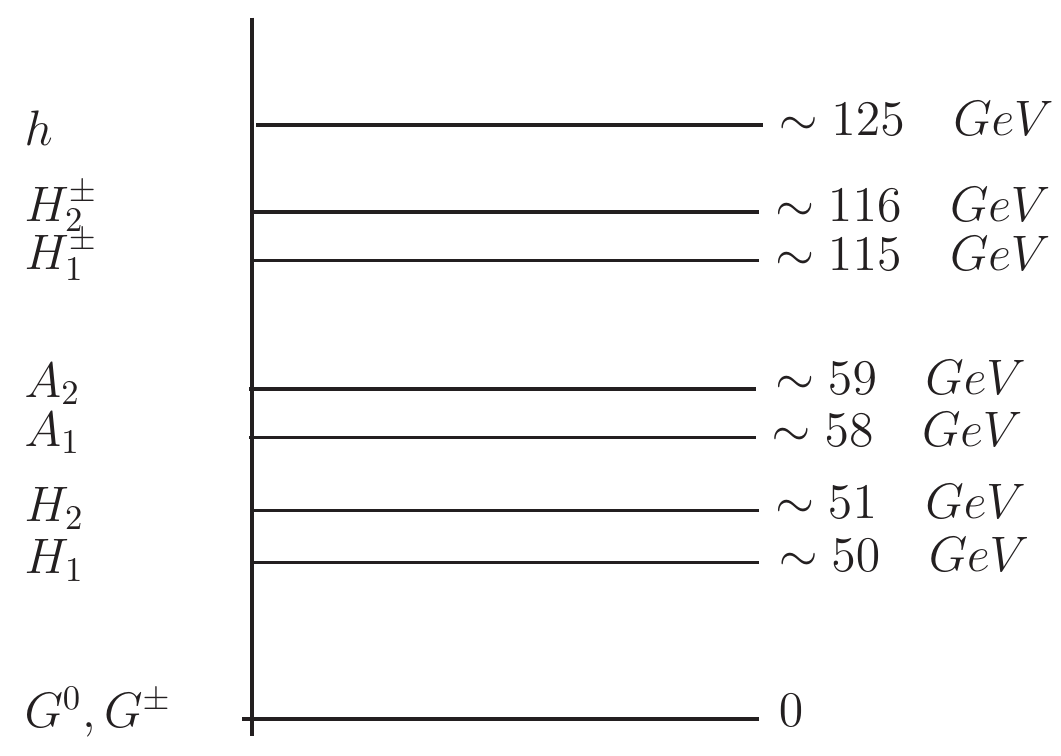}} \\
\caption{Case D. (a) Relic density vs. DM-Higgs coupling $g_{H_1H_1h}$ is shown. Here, $m_{H_1}$ changes from $45$ to $59$ GeV. Horizontal lines denote the Planck value $\Omega_{DM} h^2 = 0.1199 \pm 3 \sigma$, the region above is excluded. (b) The schematic mass spectrum for which a mid value of $m_{H_1}$ has been chosen and masses are roughly approximated.} 
\end{figure} 

\end{itemize}

The benchmark points studied above represent a typical behaviour in this region of DM mass. Clearly, as we have shown, the mass splitting between a  DM candidate and other $Z_2$-odd particles influences the freeze-out mechanism and the final value of DM relic density. 

Fig.(\ref{Mass-coupling}) shows the allowed $g_{H_1H_1h}$ coupling in different mass regions, where the grey area inside the red (case A) and  blue (case D) curves are excluded by relic density data. The white region outside the curves represents smaller relic density abundance than the observed value. It is easy to see that, apart from the Higgs resonance region, couplings that lead to the proper value of relic density are much smaller in case D than they are in case A for the same values of masses.


\begin{figure}[h!]
\centering
\includegraphics[width=0.65\textwidth]{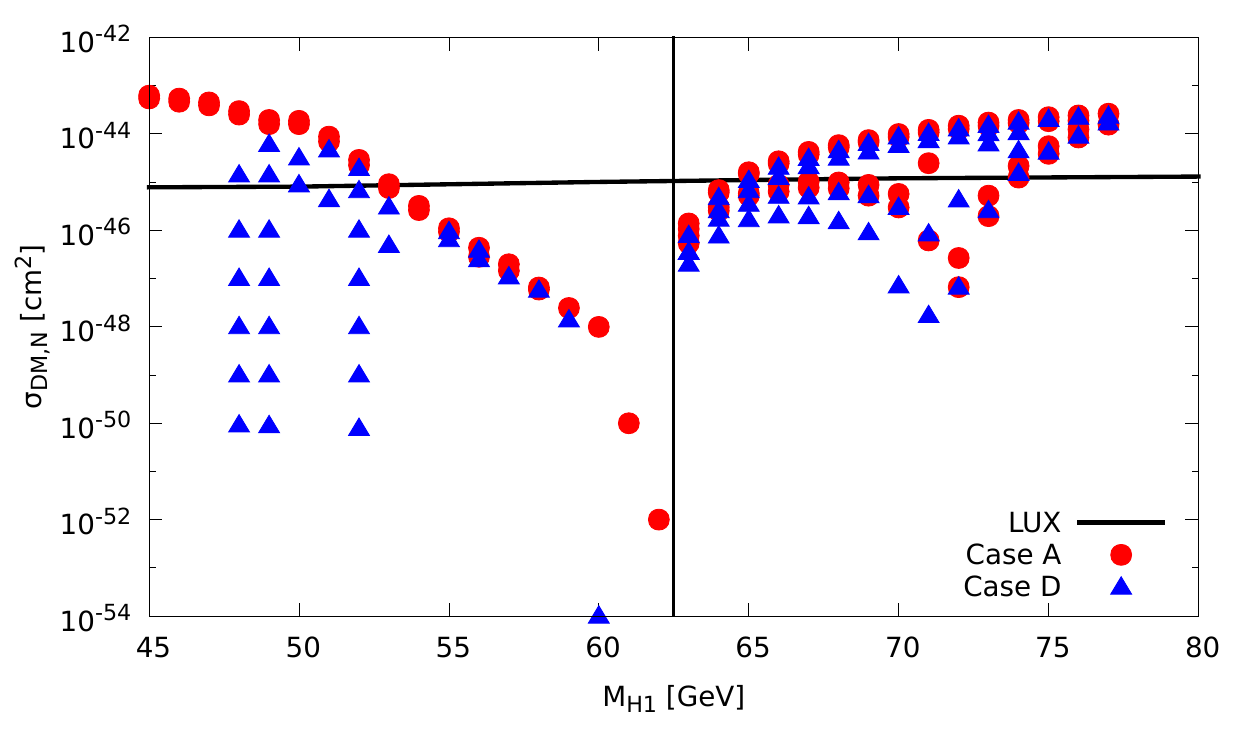}
\caption{Direct detection limits in cases A and D, where the points above the  LUX limit (horizontal black line) are excluded. The vertical line represents the Higgs-resonance region.}
\label{direct-detection-graph}
\end{figure}

\noindent
\textbf{Direct detection limits for $m_{H_1}<m_h/2$}\\
Fig.(\ref{direct-detection-graph}) compares the direct detection limits in cases A and D, where the points
above the LUX limit (black line) are excluded. The vertical line at 62.5 GeV represents the Higgs resonance mass region. 
%
For masses below $m_h/2$ the direct detection limits constrain case A much more severely than they limit case D. Masses below $53$~GeV in case A, corresponding to large $g_{H_1H_1h}$ (see Fig.(\ref{Mass-coupling})), are completely excluded in case A and only points around the Higgs-resonance region (denoted by the black vertical line) are allowed.
In case D, however, almost the whole mass region (in the $m_{H_1}<m_h/2$ range) surviving the relic density constraints agrees with the direct detection limits. 

Around the Higgs-resonance region very small $g_{H_1H_1h} \approx 10^{-3}$ couplings lead to the $\Omega_{DM}h^2$ in agreement with Planck. This corresponds to a very small DM-nucleon cross section, $\sigma_{DM-N} \approx 10^{-50}$ cm$^2$ and makes the direct detection experiments impractical due to coherent neutrino scattering.


Similar to other Higgs-portal DM models, in the small $g_{H_1H_1h}$ region, loop effect contributions are in principle important and one has to move beyond tree-level calculations for more detailed descriptions, both in the relic density estimates  and the scattering cross section analysis.\\[0.5cm]
\noindent
\textbf{Higgs invisible decays}\\
Constraints arising from limits on Higgs invisible decays can easily be estimated with making some assumptions; Firstly, the Higgs-decay channels in the I(2+1)HDM are similar to the ones in the SM (in particular, contributions to the $h\to \gamma \gamma$  are negligible). Secondly, the total decay width in the I(2+1)HDM is changed with respect to the SM only through the invisible decays. Under these assumptions the Higgs invisible $Br$ is given by the ratio of the decay widths:
\be
Br(h\to inv.) \approx \sum_{i,j} \Gamma(h\to S_i S_j)/(\Gamma_h^{\rm SM}+\sum_{i,j} \Gamma(h \to S_i S_j)).
\ee
The sum runs over particles $S_i$ of masses smaller than $m_h/2$, meaning that in the I(2+1)HDM there can be up to four particles contributing to $Br(h\to inv.)$\footnote{Recall that the SM-like (active) Higgs cannot decay invisibly into pairs of the charged scalars because of the LEP limits on their masses.}. 

\begin{figure}[ht!]
\centering
\subfloat[]{\label{invcompA}\includegraphics[width=0.45\textwidth]{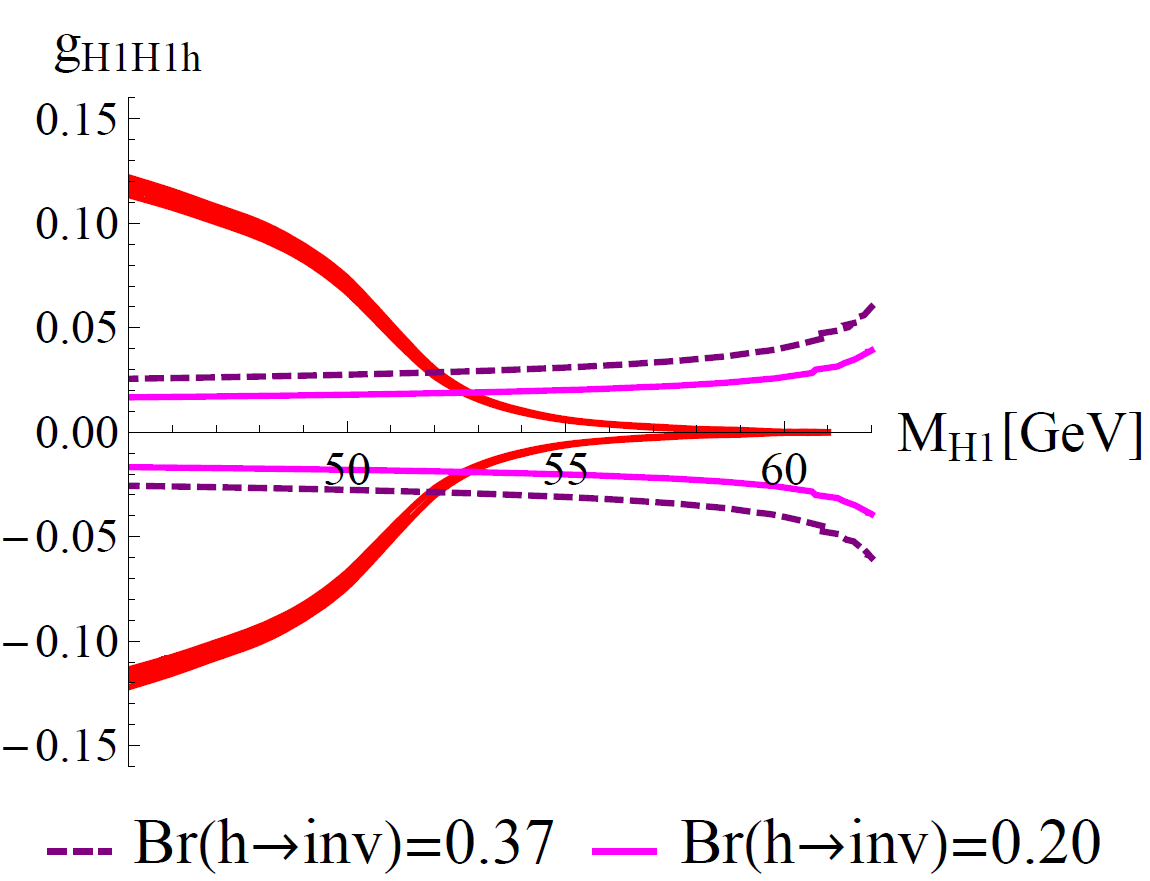}} \quad
\subfloat[]{\label{invcompD}\includegraphics[width=0.45\textwidth]{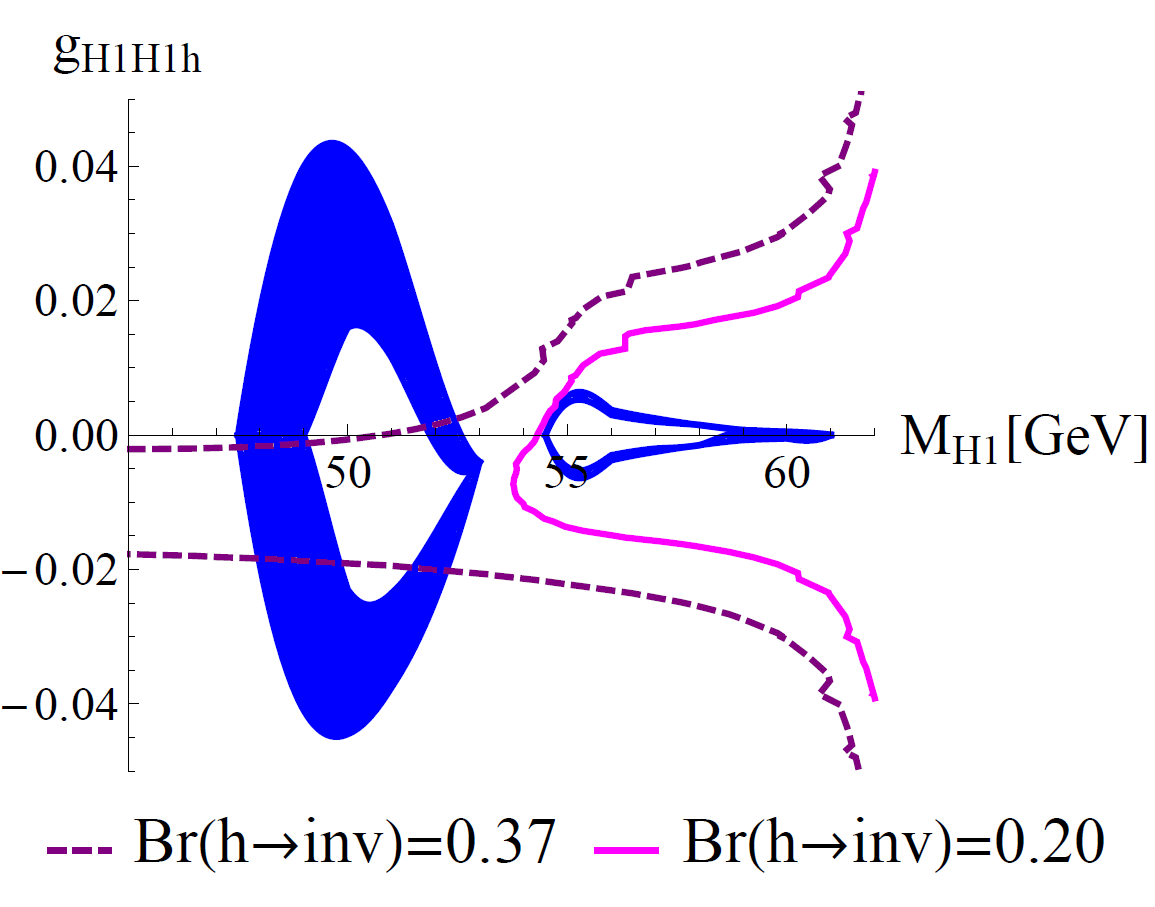}} \\
\caption{The $m_{\rm DM}$ vs. $g_{H_1H_1h}$ plane in presence of limits on the 
 Higgs invisible decay rates combined with relic density measurements for (a) case A and (b) case D.} 
\label{scanbr}
\end{figure}

If only one particle ($H_1$) is lighter than $m_h/2$ (Case A, Fig.(\ref{invcompA})) then the constraints from the ATLAS limit $Br(h\to inv) < 37\%$ \cite{atlasbr} are similar to those obtained for other Higgs-portal DM models, such as the
I(1+1)HDM. An allowed value for the Higgs-DM coupling is roughly $|g_{H_1H_1h}| \lesssim 0.002$. Using the global fit value, slightly reduces the allowed value to  $|g_{H_1H_1h}| \lesssim 0.0015$.  

For case D, where other neutral scalars can  also contribute to the Higgs invisible decays, obtained constraints are more severe, see Fig.(\ref{invcompD}). For the current experimental value $Br(h\to inv) < 37\%$ allowed values of the coupling are $-0.0015\lesssim g_{H_1H_1h} \lesssim 0$ for masses $m_{H_1} \lesssim 50$ GeV and $|g_{H_1H_1h}| < 0.02$ for larger masses.  However, demanding that $Br(h\to inv) < 20\%$ excludes all masses below  $m_{H_1} \lesssim 53 \; \mbox{GeV}$, independently of the value of the coupling. 

It is interesting to compare these limits, firstly, with the regions of proper relic density and, secondly,  with exclusion limits obtained from the direct detection experiments. From Fig.(\ref{invcompA}) it is clear that, for case A, it is not possible to fulfil Planck measurements and LHC measurements for masses $m_{H_1} \lesssim 53 \; \mbox{GeV}$, as the region in agreement with LHC corresponds to having too much DM in the Universe. These limits are comparable with those provided by the LUX experiment and delineating a region which fulfils both requirements is related to entering the Higgs resonance region, where a very small coupling still results in the good relic density, without violating LHC or direct detection bounds. 
The situation is different in case D, see Fig.(\ref{invcompD}), which in general is not constrained by direct detection limits. Here we can see that the LHC results provide severe constraints; while it is possible to fulfil $Br(h\to inv) < 37\%$ for masses below $m_h/2$ (such limits again are comparable to those provided by LUX), using the global fit value for $Br(h\to inv)$ excludes DM candidates with masses $m_{H_1} \lesssim 53 \; \mbox{GeV}$, just like in case A.

To summarize, cases A and D depend on different mechanisms to obtain proper relic density and therefore are differently constrained by direct detection experiments. However, LHC limits constrain them equally, leaving only $m_{H_1} \gtrsim 53 \; \mbox{GeV}$.


\subsubsection{Closed invisible channels ($m_W>m_{\rm DM} > m_h/2$)}

In this mass region the DM phenomenology is heavily influenced by the annihilation into gauge bosons, which leads to a different behaviour compare to the $m_{H_1} < m_h/2$ region .

The relic density values are dominated by three couplings, $g_{DMVV}$, $g_{hVV}$ and $g_{H_1H_1h}$. The first two couplings are set by gauge interactions, therefore, the behaviour of the relic density plots are ruled by the value of $g_{H_1H_1h}$. 
Since this type of annihilation is given by the strength of gauge couplings and therefore is usually very effective, proper relic density is obtained due to the cancellation effects between $H_1H_1 \to VV$ and $H_1H_1 \to h \to VV$ channels. 
For $g_{H_1H_1h}>0$ the annihilation cross section is large leading to small relic density values. As $g_{H_1H_1h}$ goes towards more negative values the annihilation cross section reduces, leading to larger values in relic density.  

Below we present in detail the numerical results obtained for case A ($\Delta, \delta_i \approx 50$ GeV) and case D ($\Delta = 1$ GeV, $\delta_A = 7$ GeV) studied in the previous section. 

\begin{itemize}
\item
\textbf{Case A}\\
Fig.(\ref{scan11a}) shows the relic density for different values of $g_{H_1H_1h}$ in case A, where different colors correspond to different DM masses. Note that the allowed relic density band is predominantly populated by negative $g_{H_1H_1h}$ values.

\begin{figure}[ht!]
\centering
\subfloat[]{\label{plotAnew}\includegraphics[width=0.45\textwidth]{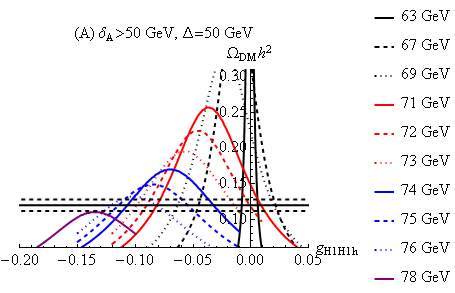}} \quad
\subfloat[]{\label{Anew-Masses}\includegraphics[width=0.4\textwidth]{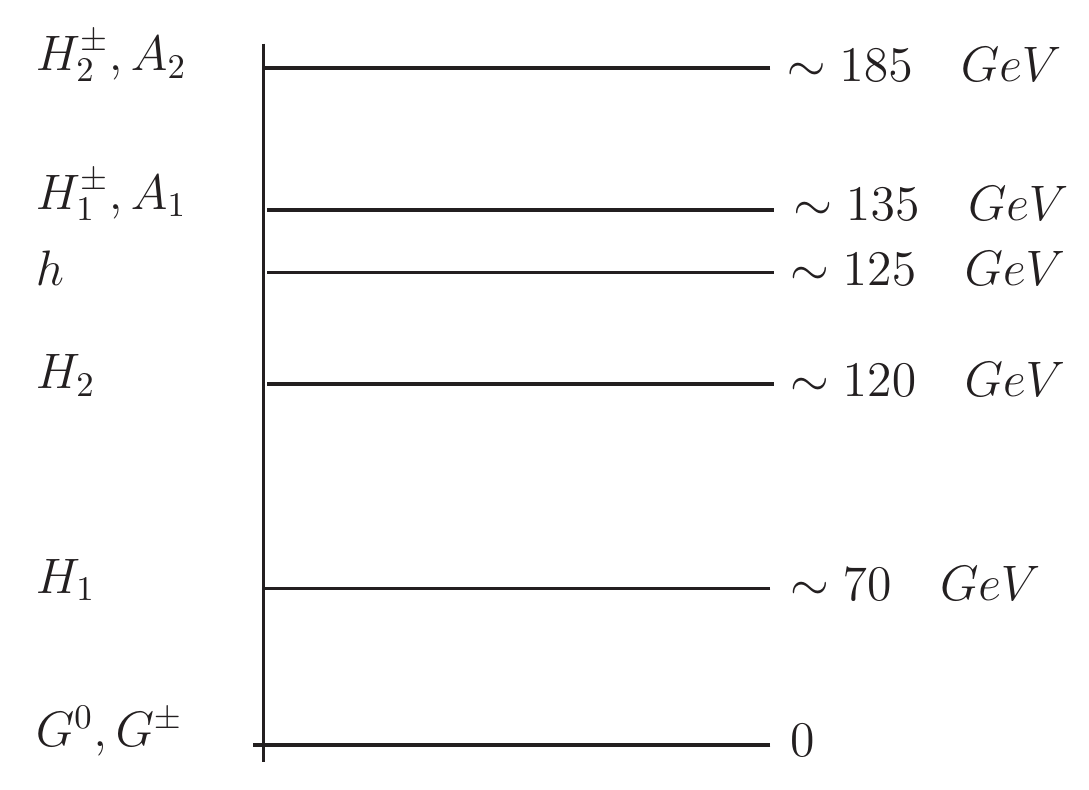}} \\
\caption{Case A. (a) Relic density vs. DM-Higgs coupling is shown. Here, $m_{H_1}$ changes from $62$ to $77$ GeV. Horizontal lines denote the Planck value $\Omega_{DM} h^2 = 0.1199 \pm 3 \sigma$, the region above is excluded. (b) Schematic mass spectrum for which a mid value of $m_{H_1}$ has been chosen and masses are roughly approximated.} 
\label{scan11a}
\end{figure} 

\item
\textbf{Case D}\\
Results for case D are presented in Fig.(\ref{scan11d}). The existence of coannihilation channels drives the relic density to smaller values in comparison to case A in Fig.(\ref{scan11a}). However, the difference is not nearly as pronounced as it was in the $m_{H_1} < m_h/2$ region.

\begin{figure}[ht!]
\centering
\subfloat[]{\label{plotDnew}\includegraphics[width=0.45\textwidth]{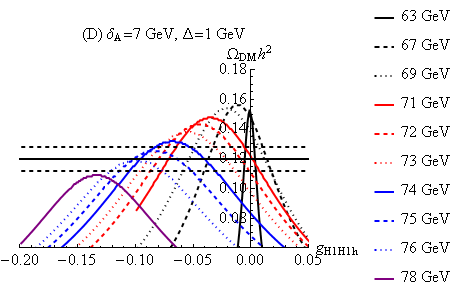}} \quad
\subfloat[]{\label{Dnew-Masses}\includegraphics[width=0.4\textwidth]{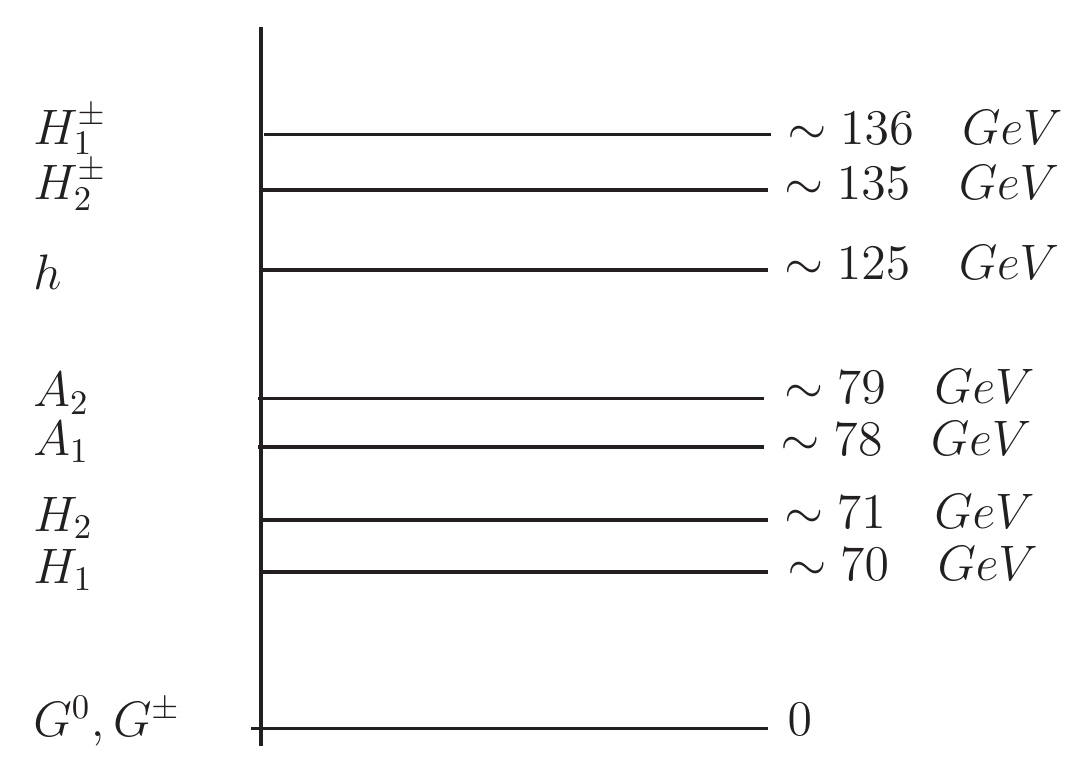}} \\
\caption{Case D. (a) Relic density vs. DM-Higgs coupling is shown. Here, $m_{H_1}$ changes from $62$ to $77$ GeV. Horizontal lines denote the Planck value $\Omega_{DM} h^2 = 0.1199 \pm 3 \sigma$, the region above is excluded. (b) Schematic mass spectrum for which a mid value of $m_{H_1}$ has been chosen and masses are roughly approximated.} 
\label{scan11d}
\end{figure} 

\end{itemize}

\begin{figure}[h!]
\centering
\includegraphics[width=0.8\textwidth]{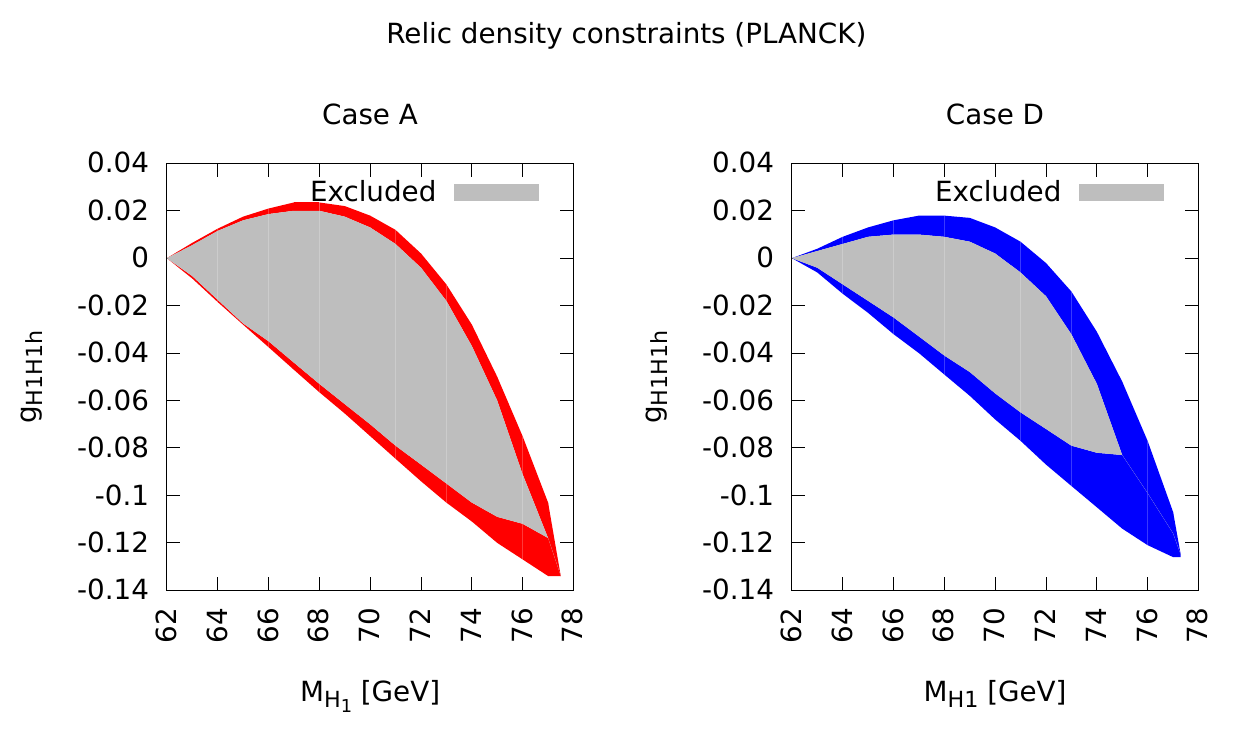} 
\caption{Relic density constraints on the mass of the DM candidate and its coupling to SM Higgs boson, with the white and gray regions representing too little and too much relic abundance respectively. Note that the regions are overlapping.}
\label{Mass-coupling2}
\end{figure}

Fig.(\ref{Mass-coupling2}) shows the allowed $g_{H_1H_1h}$ coupling in different mass regions, where the grey area inside the red (case A) and blue (case D) curves are excluded by relic density data. The white region outside the curves represents a relic density abundance smaller than the observed value. 
It is clear that the two scenarios correspond to  very similar values of coupling and the mass splitting does not play as important a role here as previously. Therefore, the direct detection exclusions will be similar in both cases (as shown in  Fig.(\ref{direct-detection-graph})), with case D being slightly less constrained than case A.

Let us finally comment briefly on the other two scenarios discussed in the previous section, namely cases B and C. As before, for case B$_8$ with $\Delta =8$ GeV, we reproduce results from case A. If $\Delta = 1$ GeV, the couplings which correspond to the proper relic density will be $30-50 \%$ larger than for the same mass in case A. This will lead to a larger DM-nucleon scattering cross section and stronger exclusion from direct detection experiments. Case C again corresponds to a subdominant DM candidate with relic density below the observed value. Detailed plots showing differences between discussed cases are presented in section \ref{summary-plots}.

\subsubsection{Heavy DM mass ($m_{\rm DM} \gg m_W$)}

As the mass splitting between inert particles is given by the quartic couplings $\lambda_2$ and $\lambda_3$, unitarity bounds for these parameters will lead to an almost degenerated mass spectrum in the heavy mass regime. Therefore, the only scenario leading to acceptable relic density values is case F, discussed in section \ref{annihilation-scenarios}, where all inert particles have similar masses. Proper relic density here is obtained through cancellations among diagrams (see Figs.(\ref{annihilation-2}),(\ref{annihilation-4}),(\ref{annihilation-5})) and a relatively large value of $g_{H_1H_1h}$ is also needed.

Similar to other multi-scalar models, this mass region escapes both the current direct detection limits and LHC constraints. However, interesting indirect detection signatures (connected to the possibility of internal bremsstrahlung $H_1 H_1 \to W^+ W^- \gamma$ processes generated through the exchange
of any of the two charged scalars $H^\pm_{1,2}$ in the $t$-channel) could arise here, which will require further studies.

\subsection{Summary of $k=1$ results for fixed DM mass}\label{summary-plots}

\begin{figure}[ht!]
\vspace{-10pt}
  \centering
  \subfloat[]{\label{m45}\includegraphics[width=0.45\textwidth]{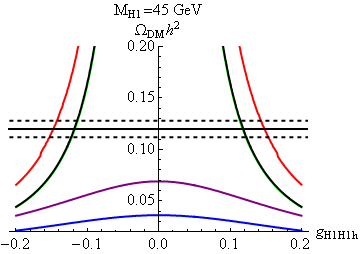}} \quad
    \subfloat[]{\label{m50}\includegraphics[width=0.45\textwidth]{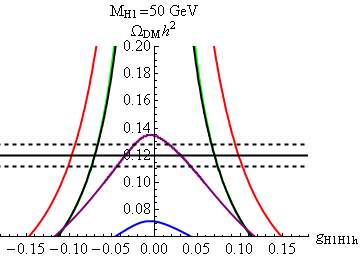}} \\
      \subfloat[]{\label{m53}\includegraphics[width=0.45\textwidth]{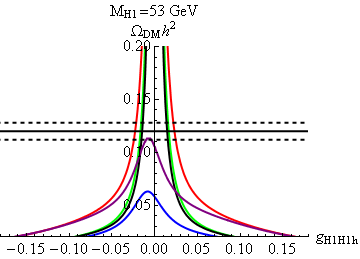}} \quad
  \subfloat[]{\label{m58}\includegraphics[width=0.45\textwidth]{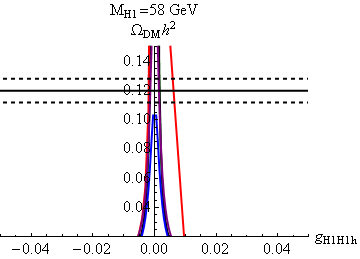}}\\
  \subfloat[]{\label{m62}\includegraphics[width=0.65\textwidth]{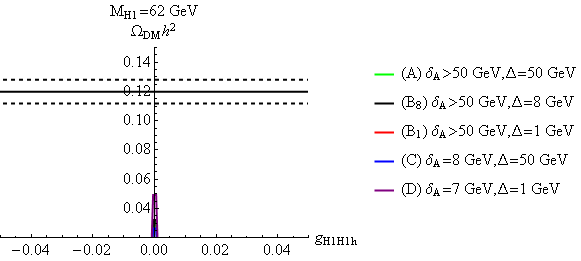}}    \\
\vspace{-5pt}
  \caption{Results for $k=1$ and fixed DM masses smaller than half the Higgs mass (a) $45$ GeV, (b) $50$ GeV, (c) $53$ GeV, (d) $58$ GeV, (e) $62$ GeV. 
  Relic density vs. DM-Higgs coupling for cases A, B$_8$, B$_1$, C and D. (Note that cases A and B$_8$ overlap.) Horizontal lines denote the Planck value $\Omega_{DM} h^2 = 0.1199 \pm 3 \sigma$, the region above is excluded. \label{scan2}}
\end{figure}

\begin{figure}[ht!]
\vspace{-10pt}
  \centering
  \subfloat[]{\label{m63}\includegraphics[width=0.45\textwidth]{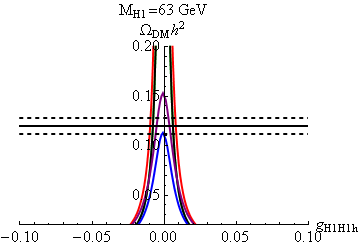}} \quad
    \subfloat[]{\label{m67}\includegraphics[width=0.45\textwidth]{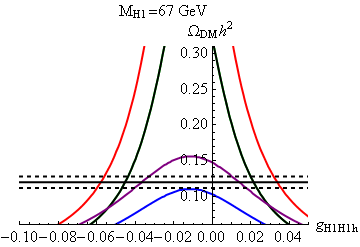}} \\
      \subfloat[]{\label{m71}\includegraphics[width=0.45\textwidth]{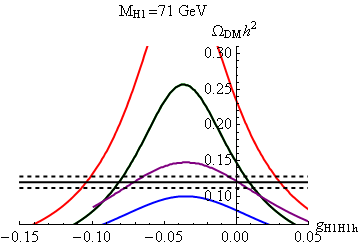}} \quad
  \subfloat[]{\label{m74}\includegraphics[width=0.45\textwidth]{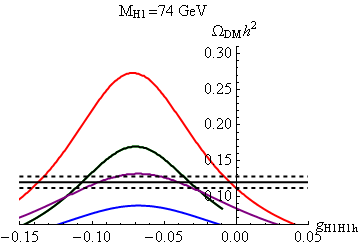}}\\
  \subfloat[]{\label{m77}\includegraphics[width=0.65\textwidth]{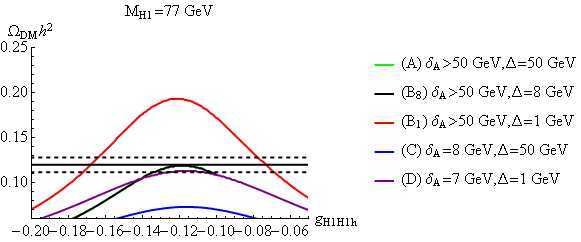}}    \\
\vspace{-5pt}
  \caption{Results for $k=1$ and fixed DM masses greater than half the Higgs mass (a) $63$ GeV, (b) $67$ GeV, (c) $71$ GeV, (d) $74$ GeV, (e) $77$ GeV. 
  Relic density vs. DM-Higgs coupling for cases A, B$_8$, B$_1$, C and D.  (Note that cases A and B$_8$ overlap.) Horizontal lines denote the Planck value $\Omega_{DM} h^2 = 0.1199 \pm 3 \sigma$, the region above is excluded. \label{scan3}}
\end{figure}

Below we present the detailed comparison between different scenarios studied in section \ref{section-keq1M} for 
$k=1$ and fixed values of DM mass. Figs.(\ref{scan2}) and (\ref{scan3}) show the relic density of the DM candidate vs. DM-Higgs coupling for DM masses less than and greater than half the Higgs mass,
respectively. 

In all plots case A (green) and case B$_8$ with $\Delta = 8$ GeV (black) are indistinguishable; coannihilation effects with $H_2$ play no role here. To show the relevance of the value of $\Delta$ we also plot case B$_1$ with $\Delta = 1$ GeV (red). It is clear that in case B$_1$ the coupling which gives $\Omega_{DM}h^2$ in agreement with Planck measurements is equal or larger than the one from case A. Therefore exclusion limits in case B$_1$ are stronger and this scenario does not provide any solution to the problems of Higgs-portal DM models. 

Case C (blue), which present an equivalent scenario to that of coannhilation in the I(1+1)HDM, is always below the Planck limit. Case D (purple) generally corresponds to couplings smaller than in cases A and B.

As a function of the DM mass, we observe the following:
\begin{itemize}
\item 
Starting from $m_{H_1} = 45$ GeV, only the cases A and B are in agreement with Planck/WMAP results. The coannihilation effects in other cases are too strong to lead to sufficient relic density values. 
\item 
From $m_{H_1} = 47$ GeV, double coannihilation effects in case D starts to appear, making this case agree with the lower Planck bounds. Notice in particular that,  without coannihilation effects the acceptable DM-Higgs coupling values are $g_{H_1H_1h} \sim 0.2$ while with coannihilation effect the DM-Higgs coupling values could be reduced to $g_{H_1H_1h}\sim 0.01$. 
\item 
As the $m_{H_1}$ grows, a smaller $g_{H_1 H_1 h}$ is required for cases A and B. Furthermore, the closer we get to the Higgs resonance, the smaller the $g_{H_1 H_1 h}$ coupling needs to be (up to an excluded region with $m_{H_1} \gtrsim 60$ GeV and a non-zero $g_{H_1H_1h}$).

\item For masses above $m_h/2$ we observe a gradual shift towards negative values of $g_{H_1H_1h}$. 
\item With $m_{H_1}$ growing towards $m_W$, the resulting relic density decreases.
\end{itemize}

\subsection{The $k\neq1$ with vanishing mixing case}\label{section-kgr1NM}
Here we study the model with $k<1$ in the no-mixing limit, $\mu^2_{12}=0$.

No mixing between the doublets, $\mu^2_{12}=0$, leads to
\be 
\tan 2\theta_h  = 0 \quad \rightarrow \theta_h = 0,\pi/2 
\ee
which in turn leads to
\bea
&& m^2_{H_1} = k (-\mu^2_2 +\Lambda_{\phi_2}) =k m^2_{H_2}, \qquad \mbox{for} \quad \theta_h =0,\\
&& m^2_{H_2} = k(-\mu^2_2 +\Lambda_{\phi_2})=k m^2_{H_1}, \qquad \mbox{for} \quad \theta_h =\pi/2. \nonumber
\eea
Note that, depending on the choice of $\theta_h$, $H_2$ or $H_1$  can play the role of the DM candidate. 

Choosing a $ \theta_h =0$ value for $k<1$ results in $H_1$ to be the DM candidate. In the $k>1$ case $H_2$ would be the DM candidate with the same choice of $\theta_h$.

The rest of the mass spectrum has the following form with $\theta_c = \theta_a =0 $: 
\bea 
&& m^2_{H^\pm_1} = k (-\mu^2_2 +\Lambda'_{\phi_2})= k m^2_{H^\pm_2}, \\
&& m^2_{A_1} = k (-\mu^2_2 +\Lambda''_{\phi_2}) = k m^2_{A_2}. \nonumber
\eea

The parameters of the model are therefore defined as follows:
\bea
&& \lambda'_{23} = \frac{1}{v^2}(m^2_{A_1}+m^2_{H_1}-2m^2_{H^\pm_1}),  \\[2mm]
&& \lambda_{2} = \frac{1}{2v^2}(m^2_{H_1}-m^2_{A_1}), \nonumber\\[2mm]
&& \lambda_{23} = g_{H_1 H_1 h} - (\frac{2}{v^2})(m^2_{H_1}-m^2_{H^\pm_1}),  \nonumber\\[2mm]
&& \mu^2_{2} = \frac{v^2}{2} g_{H_1 H_1 h} -m^2_{H_1} \nonumber
\eea
and are equivalent to those obtained from the $k=1$ with $\mu_{12}^2\neq0$ case. Again, we can perform the analysis by using the physical masses and $g_{H_1H_1h}$ coupling.

The surviving relevant non-zero Feynman rules  are the following:
\bea
&& H^+_1 H^-_1 \longrightarrow h \qquad \qquad k\lambda_{23}  v \nonumber\\
&& H^+_2 H^-_2 \longrightarrow h \qquad \qquad  \lambda_{23}  v \nonumber\\
&& H_1 H_1 \longrightarrow h \qquad \qquad \frac{k}{2}(\lambda_{23}+ \lambda'_{23} +2\lambda_{2}) v \nonumber\\
&& H_2 H_2 \longrightarrow h \qquad \qquad \frac{1}{2}(\lambda_{23}+ \lambda'_{23} +2\lambda_{2}) v \nonumber\\
&& A_1 A_1 \longrightarrow h \qquad \qquad \frac{k}{2}(\lambda_{23}+ \lambda'_{23} -2\lambda_{2}) v \nonumber\\
&& A_2 A_2 \longrightarrow h \qquad \qquad \frac{1}{2}(\lambda_{23}+ \lambda'_{23} -2\lambda_{2}) v \nonumber\\
&& H_2 A_2, H_1 A_1 \longrightarrow Z \qquad \qquad   \frac{1}{2}(g\cos\theta_W + g'\sin\theta_W) (K+K')^\mu \nonumber\\
&& H^+_1 H^-_1, H^+_2 H^-_2 \longrightarrow \gamma \qquad \qquad \frac{i}{2}(g\sin\theta_W + g'\cos\theta_W)(K+K')^\mu \nonumber\\
&& H^+_1 H^-_1, H^+_2 H^-_2 \longrightarrow Z \qquad \qquad \frac{i}{2}(g\cos\theta_W - g'\sin\theta_W)(K+K')^\mu \nonumber\\
&& H^\pm_2 H_2, H^\pm_1 H_1 \longrightarrow W^\pm \qquad \qquad \frac{ig}{2} (K+K')^\mu \nonumber\\
&& H^\pm_2 A_2, H^\pm_1 A_1 \longrightarrow W^\pm \qquad \qquad  \frac{g}{2} (K+K')^\mu \nonumber
\eea

It is possible to obtain all scenarios discussed in section \ref{section-keq1M} here. Cases A and C will correspond to small values of $k$, of the order $0.25$, while cases B and D correspond to $k$ values close to $1$. The parameters of the potential depend on masses and couplings in the same way, therefore, we conclude that cases in the $k = 1$ scenario (with mixing between inert doublets) have similar DM phenomenology to cases in the $k<1$ scenario with no mixing.

\subsection{The $k\neq1$ with small non-vanishing mixing case}\label{section-kgr1M}
In this  limit, $\theta_h$ is small and we could use approximate values for sine and cosine functions:
\be   
\sin\theta_i \approx \theta_i , \quad \cos\theta_i \approx 1.
\ee 
The mass (squared) of the DM candidate is then
\bea
m^2_{H_1} &=& (-\mu^2_2 +\Lambda_{\phi_2}) \biggl[k(1-\cdots)^2 + (\theta_h - \cdots)^2 \biggr] - \mu^2_{12} \biggr[2\theta_h - \cdots \biggr] \\
&=& (-\mu^2_2 +\Lambda_{\phi_2})k - 2\theta_h \mu^2_{12}.  \nonumber
\eea

Keeping only up to first order in $\theta_i$, the inert mass spectrum has the following form:
\bea
\label{kN1-MN1-masses}
&& m^2_{H_1} = (-\mu^2_2 +\Lambda_{\phi_2})k  - 2\theta_h\mu^2_{12} , \qquad m^2_{H_2} = (-\mu^2_2 +\Lambda_{\phi_2})   + 2\theta_h\mu^2_{12},  \\
&& m^2_{H^\pm_1} =  (-\mu^2_2 +\Lambda'_{\phi_2})k  - 2\theta_c\mu^2_{12}, \qquad m^2_{H^\pm_2} =  (-\mu^2_2 +\Lambda'_{\phi_2})  + 2\theta_c\mu^2_{12}, \nonumber\\
&& m^2_{A_1} = (-\mu^2_2 +\Lambda''_{\phi_2})k  - 2\theta_a\mu^2_{12} , \qquad m^2_{A_2} = (-\mu^2_2 +\Lambda''_{\phi_2})  + 2\theta_a\mu^2_{12}. \nonumber
\eea

The parameters of the model are therefore defined as follows:
\bea
&& \lambda'_{23} = \frac{1}{kv^2}\biggl[ m^2_{H_1} + m^2_{A_1} - 2 m^2_{H^\pm_1} + 2\mu^2_{12} (\theta_h +\theta_a -2\theta_c) \biggr],  \nonumber\\
&& \lambda_{2} = \frac{1}{2kv^2}\biggl[m^2_{H_1} - m^2_{A_2} + 2\mu^2_{12} (\theta_h -\theta_a)  \biggr], \nonumber\\
&& \lambda_{23} = \frac{2}{kv^2}\biggl[m^2_{H^\pm_1} - m^2_{H_1} + 2\mu^2_{12} (\theta_c -\theta_h)\biggr] + g_{H_1 H_1 h},  \nonumber\\
&& \mu^2_{2} = \frac{v^2}{2} g_{H_1 H_1 h} - \frac{1}{k}(m^2_{H_1} + 2\mu^2_{12} \theta_h),  \nonumber\\
&& \theta_h = \frac{-m^2_{H_1}}{\mu^2_{12}} \pm \sqrt{ \frac{m^4_{H_1}}{\mu^4_{12}} - \frac{8k}{k-1}}, \label{kN1-MN1-angles} \\
&& \theta_c = \frac{-m^2_{H^\pm_1}}{\mu^2_{12}} \pm \sqrt{ \frac{m^4_{H^\pm_1}}{\mu^4_{12}} - \frac{8k}{k-1}}, \nonumber \\
&& \theta_a = \frac{-m^2_{A_1}}{\mu^2_{12}} \pm \sqrt{ \frac{m^4_{A_1}}{\mu^4_{12}} - \frac{8k}{k-1}}. \nonumber  
\eea

Then, the surviving relevant Feynman rules in this case are the following:
\bea
&& H^+_1 H^-_1 \longrightarrow h  \qquad \lambda_{23}  v\nonumber\\
&& H^+_2 H^-_2 \longrightarrow h  \qquad k\lambda_{23} v \nonumber\\ 
&& H^\pm_2 H^\mp_1 \longrightarrow h  \qquad  \lambda_{23}(1-k)\theta_c  v \nonumber\\ 
&& H_1 H_1 \longrightarrow h \qquad  (\lambda_{23}+ \lambda'_{23}+2\lambda_{2}) \frac{v}{2}\nonumber\\ 
&& H_2 H_2 \longrightarrow h \qquad  (\lambda_{23}+ \lambda'_{23}+2\lambda_{2})k\frac{v}{2} \nonumber\\ 
&& H_2 H_1 \longrightarrow h  \qquad (\lambda_{23}+\lambda'_{23}+2\lambda_{2} )(1-k)  \theta_h  v \nonumber\\ 
&& A_1 A_1 \longrightarrow h  \qquad (\lambda_{23}+ \lambda'_{23}-2\lambda_{2}) \frac{v}{2}\nonumber\\ 
&& A_2 A_2 \longrightarrow h  \qquad (\lambda_{23}+ \lambda'_{23}-2\lambda_{2})k \frac{v}{2}\nonumber\\ 
&& A_2 A_1 \longrightarrow h  \qquad (\lambda_{23}+\lambda'_{23}+2\lambda_{2} )(1-k) \theta_a v \nonumber\\
&& H^+_1 H^-_1, H^+_2 H^-_2 \longrightarrow \gamma \qquad \qquad \frac{i}{2}(g\sin\theta_W + g'\cos\theta_W)(K+K')^\mu \nonumber\\ 
&& H^+_1 H^-_1, H^+_2 H^-_2 \longrightarrow Z \qquad \qquad \frac{i}{2}(g\cos\theta_W - g'\sin\theta_W)(K+K')^\mu \nonumber\\ 
&& H^\pm_2 H_2, H^\pm_1 H_1 \longrightarrow W^\pm \qquad \qquad \frac{ig}{2}(K+K')^\mu \nonumber\\ 
&& H^\pm_2 H_1, H^\pm_1 H_2 \longrightarrow W^\pm \qquad \qquad \frac{ig}{2} (\theta_h -\theta_c)(K+K')^\mu \nonumber\\ 
&& H^\pm_2 A_2, H^\pm_1 A_1 \longrightarrow W^\pm \qquad \qquad  \frac{g}{2}(K+K')^\mu \nonumber\\ 
&& H^\pm_2 A_1, H^\pm_1 A_2 \longrightarrow W^\pm \qquad \qquad  \frac{g}{2} (\theta_a -\theta_c)(K+K')^\mu \nonumber\\ 
&& H_2 A_2, H_1 A_1 \longrightarrow Z \qquad \qquad   \frac{1}{2}(g\cos\theta_W + g'\sin\theta_W)(K+K')^\mu \nonumber\\ 
&& H_2 A_1, H_1 A_2 \longrightarrow Z \qquad \qquad  \frac{1}{2}(g\cos\theta_W + g'\sin\theta_W)(\theta_h -\theta_a)(K+K')^\mu \nonumber
\eea

Note that in this case the DM candidate $H_1$, annihilates faster than the heavier counterpart $H_2$ ($g_{H_2 H_2 h} = k g_{H_1 H_1 h}$ with $k<1$)\footnote{For  $k>1$ when $H_2$ becomes the lighter particle and therefore the DM candidate, $g_{H_2 H_2 h} = k g_{H_1 H_1 h}$  leads to the DM annihilating faster than $H_1$, as expected.}.


One should also note that with $k\neq1$, there exists a non-vanishing vertex $H_1 H_2 \to h$, which in principle will influence the coannihilation options in cases B and D with respect to the $k=1$ scenario studied in section \ref{section-keq1M}. However, since this coupling is proportional to $\theta_h$, and thus is small by definition, it will not introduce drastic changes to the obtained results.

\section{Conclusions}\label{conclusion}
We have studied a model with Two Inert Doublets plus One Higgs Doublet (I(2+1)HDM). The two inert doublets are $Z_2$-odd and the active doublet is $Z_2$-even which plays the role of the SM Higgs doublet. The I(2+1)HDM may be regarded as an extension to the I(1+1)HDM (also known as the Inert Doublet Model (IDM)). 

The I(2+1)HDM contains 4 neutral and 4 charged inert particles, which is double the particle content of the I(1+1)HDM. The lightest particle amongst the inert ones, stabilised by an exact $Z_2$ symmetry, provides a viable DM candidate. We have then studied the DM phenomenology in different parameter scenarios of this model. 

Similar to I(1+1)HDM, the I(2+1)HDM contains all features of a Higgs-portal scalar DM model, with certain modifications. In such models, the Higgs-DM coupling, $g_{H_1H_1h}$, governs the DM annihilation rate $\langle\sigma v\rangle$, the DM-nucleon scattering cross section $\sigma_{DM-N}$ and the Higgs invisible decays. This is the scenario studied in case A, which is the closest to the pure Higgs-portal DM model. Imposing current Planck, LUX and LHC constraints we found that, as in the I(1+1)HDM, it is not possible to have a DM candidate with mass below $53$ GeV. 

The tension in simultaneously fulfilling current experimental constraints on the processes described above can be lifted by introducing coannihilation processes offered by the rich inert scalar sector. In such case the relation between $\Omega_{DM}h^2$ and $\sigma_{DM-N}$ is destroyed, which is the scenario studied in case D. These coannihilation processes allow for sufficient relic density values in different DM mass regions compared to the I(1+1)HDM, where, for $m_{\rm DM} < m_W$, there is only a possibility of destructive coannihilation with a pseudoscalar inert particle. Constructive coannhilation with the remaining charged scalar is not possible in this region, as sufficiently light charged scalars are excluded by LEP measurements. We found that in the I(2+1)HDM coannihilation between all neutral particles, denoted as case D, can lead to a proper relic density for masses below $m_W$. Furthermore, this region, especially for masses $m_{H_1} \lesssim 50$ GeV, is in agreement with the current direct detection limits provided by the LUX experiment.

This new region of masses can survive also the current direct limits on Higgs invisible decays provided by the ATLAS and CMS collaborations at the LHC, i.e. $37\%$. However, we found that it is not possible to be in agreement with the global fit value of $Br(h\to inv) < 20 \%$, unless the DM mass is above $53$ GeV. For this model then, the LHC limits are stronger than those provided by direct detection experiments.

In heavier DM mass regions ($m_W>m_{\rm DM} > m_h/2$), where annihilation through gauge bosons contributions is dominant and reduces the relic density values, the $g_{H_1H_1h}$ coupling is driven towards negative values. The coannihilation effects are not as visible in the $m_{\rm DM} > m_h/2$ region since the contribution from gauge boson annihilation plays an important role in this region. This contribution is more pronounced closer to $m_W$ threshold, where case D corresponds to $\sigma_{DM-N}$ of an order of magnitude smaller than case A. 

All studied cases, as the general Higgs-portal models, are in agreement with the current experimental constraints in the Higgs resonance region ($\sim m_h/2$). Here, very small $g_{H_1H_1h}$ couplings are required to  fulfil relic density constraints. Such  small $g_{H_1H_1h}$ values  lead to very small DM-nucleon cross section, which makes the direct detection experiments impractical in this region, especially since it reaches the boundary of coherent neutrino scattering which has then to be taken into account. Also, in such small $g_{H_1H_1h}$ region, loop-effect contributions are important and one has to move beyond tree-level calculations for more detailed descriptions, which is the case in all Higgs portal models.

It is also interesting to note that the strong limits on the scattering through $Z$-exchange by direct detection experiments do not apply to this model since
a natural mass splitting between the scalar and pseudoscalar particles is provided by non-zero $\lambda_2$ and $\lambda_3$ parameters in the potential.

We would also like to comment here on the special case C studied in section (\ref{section-keq1M}), which always results in relic density values below the observed amount and therefore cannot account for the whole DM in the Universe. It could however provide a subdominant DM candidate in a multi-component DM model. One of the examples of such model is the I(4+2)HDM \cite{Keus:2014isa} in which the scalar sector consists of two copies of the scalar sector of our I(2+1)HDM. Case C could then become a viable scenario with sufficient relic density values.

\section*{Note Added}
As this paper was being finalised a related paper appeared which also analyses DM
in a model with `Two Inert Doublets plus One Higgs Doublet' \cite{Fortes:2014dca}.

\section*{Acknowledgements}
SM is financed in part through the NExT Institute and from the STFC Consolidated ST/ J000396/1. SFK also acknowledges partial support from the STFC Consolidated ST/J000396/ 1 and EU ITN grant INVISIBLES 289442. DS is grateful for useful comments and discussions with Maria Krawczyk and Bogumila Swiezewska. DS is financed in part by the grant NCN OPUS 2012/05/B/ST2/03306 (2012-2016). Part of VK's research was financed through a Visiting Fellowship from The Leverhulme Trust (London, UK).

\appendix

\section{Feynman rules}\label{Feyn-rules}

Here, we recap the Feynman rules of the model considered here in its most general setup.

\subsection{Scalar couplings}

These are as follows:
\bea
&& hh \longrightarrow h \qquad \qquad  \lambda_{33} v \nonumber\\ 
&& H^+_1 H^-_1 \longrightarrow h  \qquad (\lambda_{23} \cos^2\theta_c +\lambda_{31}\sin^2\theta_c) v=  \lambda_{23}( \cos^2\theta_c +k\sin^2\theta_c) v\nonumber\\
&& H^+_2 H^-_2 \longrightarrow h  \qquad (\lambda_{23}\sin^2\theta_c+\lambda_{31}\cos^2\theta_c) v = \lambda_{23}(\sin^2\theta_c+k\cos^2\theta_c) v \nonumber\\ 
&& H^\pm_2 H^\mp_1 \longrightarrow h  \qquad (\lambda_{23}-\lambda_{31})\sin\theta_c \cos\theta_c v = \lambda_{23}(1-k)\sin\theta_c \cos\theta_c v \nonumber\\ 
&& H_1 H_1 \longrightarrow h \qquad  ((\lambda_{23}+ \lambda'_{23}+2\lambda_{2})\cos^2\theta_h +( \lambda_{31} +\lambda'_{31}+2\lambda_{3})\sin^2\theta_h) \frac{v}{2} =\nonumber\\
&& \qquad \qquad \qquad \quad  (\lambda_{23}+ \lambda'_{23}+2\lambda_{2}) (\cos^2\theta_h +k\sin^2\theta_h) \frac{v}{2}\nonumber\\ 
&& H_2 H_2 \longrightarrow h \qquad  ((\lambda_{23}+ \lambda'_{23}+2\lambda_{2})\sin^2\theta_h  +( \lambda_{31} +\lambda'_{31}+2\lambda_{3})\cos^2\theta_h)\frac{v}{2}= \nonumber\\
&& \qquad \qquad \qquad \quad  (\lambda_{23}+ \lambda'_{23}+2\lambda_{2})(\sin^2\theta_h  +k\cos^2\theta_h)\frac{v}{2} \nonumber\\ 
&& H_2 H_1 \longrightarrow h  \qquad (\lambda_{23}+\lambda'_{23}+2\lambda_{2}  -(\lambda_{31}+\lambda'_{31}+2\lambda_{3})) \sin\theta_h \cos\theta_h v =\nonumber\\
&& \qquad \qquad \qquad \quad (\lambda_{23}+\lambda'_{23}+2\lambda_{2} )(1-k)  \sin\theta_h \cos\theta_h v \nonumber\\ 
&& A_1 A_1 \longrightarrow h  \qquad ((\lambda_{23}+ \lambda'_{23}-2\lambda_{2})\cos^2\theta_a +( \lambda_{31} +\lambda'_{31}-2\lambda_{3})\sin^2\theta_a) \frac{v}{2} = \nonumber\\
&& \qquad \qquad \qquad \quad  (\lambda_{23}+ \lambda'_{23}-2\lambda_{2})(\cos^2\theta_a +k\sin^2\theta_a) \frac{v}{2}\nonumber\\ 
&& A_2 A_2 \longrightarrow h  \qquad ((\lambda_{23}+ \lambda'_{23}-2\lambda_{2})\sin^2\theta_a  +( \lambda_{31} +\lambda'_{31}-2\lambda_{3})\cos^2\theta_a) \frac{v}{2} =\nonumber\\
&& \qquad \qquad \qquad \quad  (\lambda_{23}+ \lambda'_{23}-2\lambda_{2})(\sin^2\theta_a  +k\cos^2\theta_a) \frac{v}{2}\nonumber\\ 
&& A_2 A_1 \longrightarrow h  \qquad (\lambda_{23}+\lambda'_{23}+2\lambda_{2}  -(\lambda_{31}+\lambda'_{31}+2\lambda_{3})) \sin\theta_a \cos\theta_a v = \nonumber\\
&& \qquad \qquad \qquad \quad (\lambda_{23}+\lambda'_{23}+2\lambda_{2} )(1-k) \sin\theta_a \cos\theta_a v \nonumber
\eea

\subsection{Gauge couplings}

These are as follows:
\bea
&& W^+ W^- \longrightarrow h \qquad \qquad \frac{g^2}{2} v \nonumber\\[2mm]
&& ZZ \longrightarrow h \qquad \qquad \frac{1}{8}(g\cos\theta_W + g'\sin\theta_W)^2 v \nonumber\\[2mm]
&& H^+_1 H^-_1, H^+_2 H^-_2 \longrightarrow \gamma \qquad \qquad \frac{i}{2}(g\sin\theta_W + g'\cos\theta_W)(K+K')^\mu \nonumber\\[2mm]
%
&& H^+_1 H^-_1, H^+_2 H^-_2 \longrightarrow Z \qquad \qquad \frac{i}{2}(g\cos\theta_W - g'\sin\theta_W)(K+K')^\mu \nonumber\\[2mm]
%
&& H^\pm_2 H_2, H^\pm_1 H_1 \longrightarrow W^\pm \qquad \qquad \frac{ig}{2} \cos(\theta_h -\theta_c)(K+K')^\mu \nonumber\\[2mm]
&& H^\pm_2 H_1, H^\pm_1 H_2 \longrightarrow W^\pm \qquad \qquad \frac{ig}{2} \sin(\theta_h -\theta_c)(K+K')^\mu \nonumber\\[2mm]
&& H^\pm_2 A_2, H^\pm_1 A_1 \longrightarrow W^\pm \qquad \qquad  \frac{g}{2} \cos(\theta_a -\theta_c)(K+K')^\mu \nonumber\\[2mm]
&& H^\pm_2 A_1, H^\pm_1 A_2 \longrightarrow W^\pm \qquad \qquad  \frac{g}{2} \sin(\theta_a -\theta_c)(K+K')^\mu \nonumber\\[2mm]
&& H_2 A_2, H_1 A_1 \longrightarrow Z \qquad \qquad   \frac{1}{2}(g\cos\theta_W + g'\sin\theta_W) \cos(\theta_h -\theta_a)(K+K')^\mu \nonumber\\[2mm]
&& H_2 A_1, H_1 A_2 \longrightarrow Z \qquad \qquad  \frac{1}{2}(g\cos\theta_W + g'\sin\theta_W) \sin(\theta_h -\theta_a)(K+K')^\mu \nonumber
\eea
where $K$ and $K'$ are the momenta of the associated particles in the decay channel.

Note that the Yukawa couplings in the I(2+1)HDM case are identical to the SM ones.

\end{document}